\documentclass[aps,prl,twocolumn, superscriptaddress, longbibliography]{revtex4-1}

\usepackage{graphicx}
\usepackage{dcolumn}
\usepackage{bm}
\usepackage{color}
\usepackage{colortbl}
\usepackage{amsmath}
\usepackage{amssymb}
\usepackage{mathrsfs}  

\usepackage{multirow}

\begin{document}

\title{Experimental demonstration of time-frequency duality of biphotons}
\author{Rui-Bo Jin}
\affiliation{Laboratory of Optical Information Technology, Wuhan Institute of Technology, Wuhan 430205, China\\}
\author{Takuma Saito}
\affiliation{The University of Electro-Communications, 1-5-1 Chofugaoka, Chofu, Tokyo, Japan}
\author{Ryosuke Shimizu}
\email{r-simizu@uec.ac.jp}
\affiliation{The University of Electro-Communications, 1-5-1 Chofugaoka, Chofu, Tokyo, Japan}

\date{\today }

\begin{abstract}
Time-frequency duality, which enables control of optical waveforms by manipulating amplitudes and phases of electromagnetic fields, plays a pivotal role in a wide range of modern optics.
The conventional one-dimensional (1D)  time-frequency duality has been successfully applied to characterize the behavior of classical light, such as ultrafast optical pulses from a laser.
However, the 1D treatment is not enough to characterize quantum mechanical correlations in the time-frequency behavior of multiple photons, such as the biphotons from parametric down conversion.  The two-dimensional treatment is essentially required, but has not been fully demonstrated yet due to the technical problem.
Here, we study the two-dimensional (2D)  time-frequency duality  duality of biphotons, by measuring two-photon distributions in both frequency and time domains.
It was found that generated biphotons satisfy the Fourier limited condition quantum mechanically, but not classically, by analyzing the time-bandwidth products in the 2D Fourier transform.
Our study provides an essential and deeper understanding of light beyond classical wave optics, and opens up new possibilities for optical synthesis in a high-dimensional frequency space in a quantum manner.
\end{abstract}



\maketitle

\begin{figure}[tbhp]
\centering
\includegraphics[width= 0.45\textwidth]{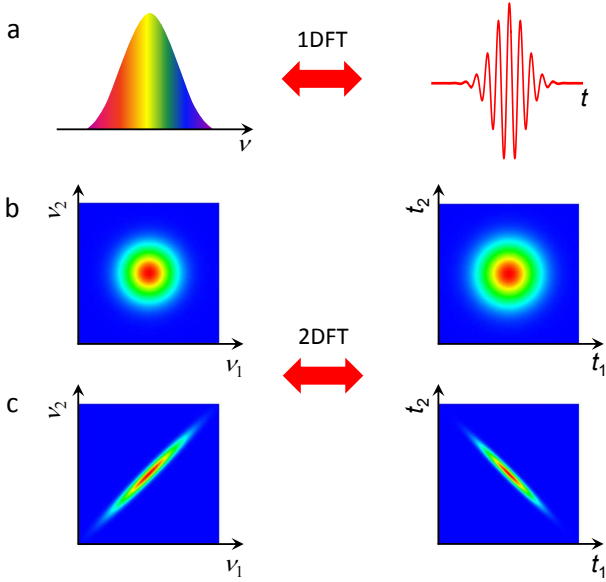}
\caption{  \textbf{Conceptual drawing of  1- and 2D treatments for the time-frequency duality.}
\textbf{(a)} Classical duality in the time-frequency domains; a broader distribution in the frequency domain results in a narrower distribution in the time domain;
\textbf{(b)} 2D Fourier transform for a frequency-uncorrelated biphoton; frequency uncorrelation leads to uncorrelation in the time domain;
\textbf{(c)} 2D Fourier transform of a biphoton with positive frequency correlation; correlations are inverted between the time and frequency domains.
}
\label{duality}
\end{figure}

\textbf{\emph{Introduction}}
Optical studies that utilize the relationship coupled by Fourier transform are called Fourier optics, and form a wide range of optical science and technology fields, including optical imaging \cite{Stark1982, Goodman2017, Zhang2014}, spectroscopy \cite{Smith2011, Robert1972}, optical measurement \cite{Sirohi2017, Griffiths2007}, and optical signal processing \cite{VanderLugt2005, Rhodes2009}.
In particular, it is well-known that each electric field distribution in the time and frequency domains is connected by the one-dimensional Fourier transform (1DFT)
\begin{equation}
\label{eq1}
A^{(1)}(t) = \frac{1}{\sqrt {2\pi }  }\int d\nu \tilde A^{(1)}(\nu )  e^{ - 2\pi i\nu \,t}
\end{equation}
where $A^{(1)}(t)$ is the electric field amplitude as a function of time $t$, and $\tilde A^{(1)}(\nu)$ is the corresponding distribution with a frequency $\nu$, as shown in Fig.\,\ref{duality}(a).
This relationship is a fundamental principle of cutting-edge technologies on  an ultrashort pulse laser, as seen in the recent developments in optical synthesis \cite{Chan2011, Suhaimi2015}.
On the other hand, Fourier optical phenomena can also be explained from the viewpoint of the particle nature of light, i.e., the photon.
Since the photons contained in a laser light pulse have no quantum correlation, the collective behavior of a many-photon system can be handled as a single-photon problem \cite{Kocsis2011, Aspden2016}.
As a result, the quantum mechanical treatment between the time and frequency domains of a laser light pulse can be explained by 1DFT, producing results in equivalent to an understanding with classical wave optics.
However, in principle, these phenomena should be treated  as a quantum many-body system, because a large number of photons are contained in an optical pulse output from an ultrashort pulse laser.

Recent progress in quantum optical technologies allows us to control not only a number of photons (i.e., photon statistics), but also the frequency quantum correlations in an optical pulse \cite{Eckstein2011, Jin2016QST, Chen2017, Jin2017PRA}.
Such a frequency quantum correlation would affect its temporal distribution directly through the time-frequency duality.
It is not reasonable to treat the behavior of a quantum-mechanically correlated photon  as a single-photon (or 1D) problem.
Therefore, an optical pulse containing the correlated photons requires a higher-order Fourier treatment that incorporates the quantum mechanics.
As the first step toward future photonics at the single-photon level in the time-frequency domain,  we focus here on the time-frequency behavior of biphotons, which requires the treatment of the two-dimensional Fourier transform (2DFT), as shown in Figs.\,\ref{duality}\textbf{b}, \textbf{c}.
\begin{equation}
\label{eq2}
A^{(2)}(t_1 ,t_2 ) = \frac{1} {2\pi} \int \int  d  \nu _1 d\nu _2 \tilde A^{(2)}(\nu _1 ,\nu _2 ) e^{ - 2\pi i(\nu _1 \,t_1  + \nu _2 \,t_2) }
\end{equation}
where $A^{(2)}(t_1 ,t_2 )$ is the biphoton probability amplitude at time $t_1$ and $t_2$, and $\tilde A^{(2)}(\nu _1 ,\nu _2 )$ is the corresponding distribution with the frequency $\nu_1$ and $\nu_2$.
%
%
%
%

From the early stage of quantum optical experiments, time-frequency correlation of biphotons generated from spontaneous parametric down-conversion process have been extensively investigated \cite{Burnham1970, Hong1987, Larchuk1993, Shih1994, Giovannetti2002, Jin2018Optica, Kim2005, Shimizu2009,  Avenhaus2009, Fang2014, Allgaier2017, Kuzucu2008PRL, Cho2014}.
In these studies, they discussed the relationship between a classical frequency spectrum and time-domain quantum interference patterns \cite{Hong1987, Larchuk1993, Shih1994, Giovannetti2002, Jin2018Optica}, the spectral properties \cite{Kim2005, Shimizu2009,  Avenhaus2009, Fang2014}, or the temporal \cite{Allgaier2017, Kuzucu2008PRL, Cho2014} properties of biphotons. These studies provide partial understanding of nonclassical behavior of biphotons.
However, it is necessary to discuss in the two-dimensional time-frequency space for the comprehensive understanding of biphoton behavior, but still has not been fully demonstrated yet due to the technical problem.
Here, we experimentally demonstrate the time-frequency duality of biphotons with the positive frequency correlations, by measuring both the two-photon spectral intensity distribution (TSI) and the two-photon temporal intensity distribution (TTI).
Furthermore, we show the variation of two-photon temporal distributions as a result of the two-photon spectral modulations, keeping quantum optical-Fourier-transform limited conditions.

\textbf{\emph{Results}}
To generate biphotons  with  positive frequency correlation, we exploit the spontaneous parametric down-conversion process in a periodically poled  potassium titanyl phosphate (PPKTP) crystal pumped by a mode-locked titanium sapphire laser, operating at the center wavelength of 792 nm.
Thanks to the group velocity matching condition \cite{Konig2004,Jin2013OE} with the femtosecond laser pulse pumping, we could generate biphotons with positive frequency correlations at the center wavelength of 1584 nm (see Supplementary Information for details).
Since biphotons generated via the type-II phase-matching condition have orthogonal polarizations, the polarization of the constituent photons were aligned along either the crystallographic y- or z- axis.
To manipulate a two-photon spectral distribution, we used  pump pulses with a bandwidth of either 8.1 or 2.8 nm.
In addition, we prepared two PPTKP crystals, one 30-mm and the other 10-mm long, and positioned one or the other in our experimental apparatus (see Methods).

We performed two experiments to characterize the biphoton distributions in the time-frequency domain.
A two-photon spectrometer consisting of two tunable bandpass filters, which had a  Gaussian-shaped filter function with a fixed FWHM of 0.56 nm, and a tunable central wavelength from 1560-1620 nm, followed by a two-photon detector enabled us to conduct the TSI measurements \cite{Shimizu2009,Jin2013OE}.
For the TTI measurements, we utilized a time-resolved upconversion detection system with a spatial multiplexing technique \cite{Kuzucu2008OL, Kuzucu2008PRL}.
The TTI was measured by scanning the temporal delay between the constituent photons and the local oscillator pulse coming from the mode-locked laser with a step length of 0.13 ps, and we recorded coincidence events (see Methods).

\begin{figure*}[tbhp]
\centering
\includegraphics[width= 0.85\textwidth]{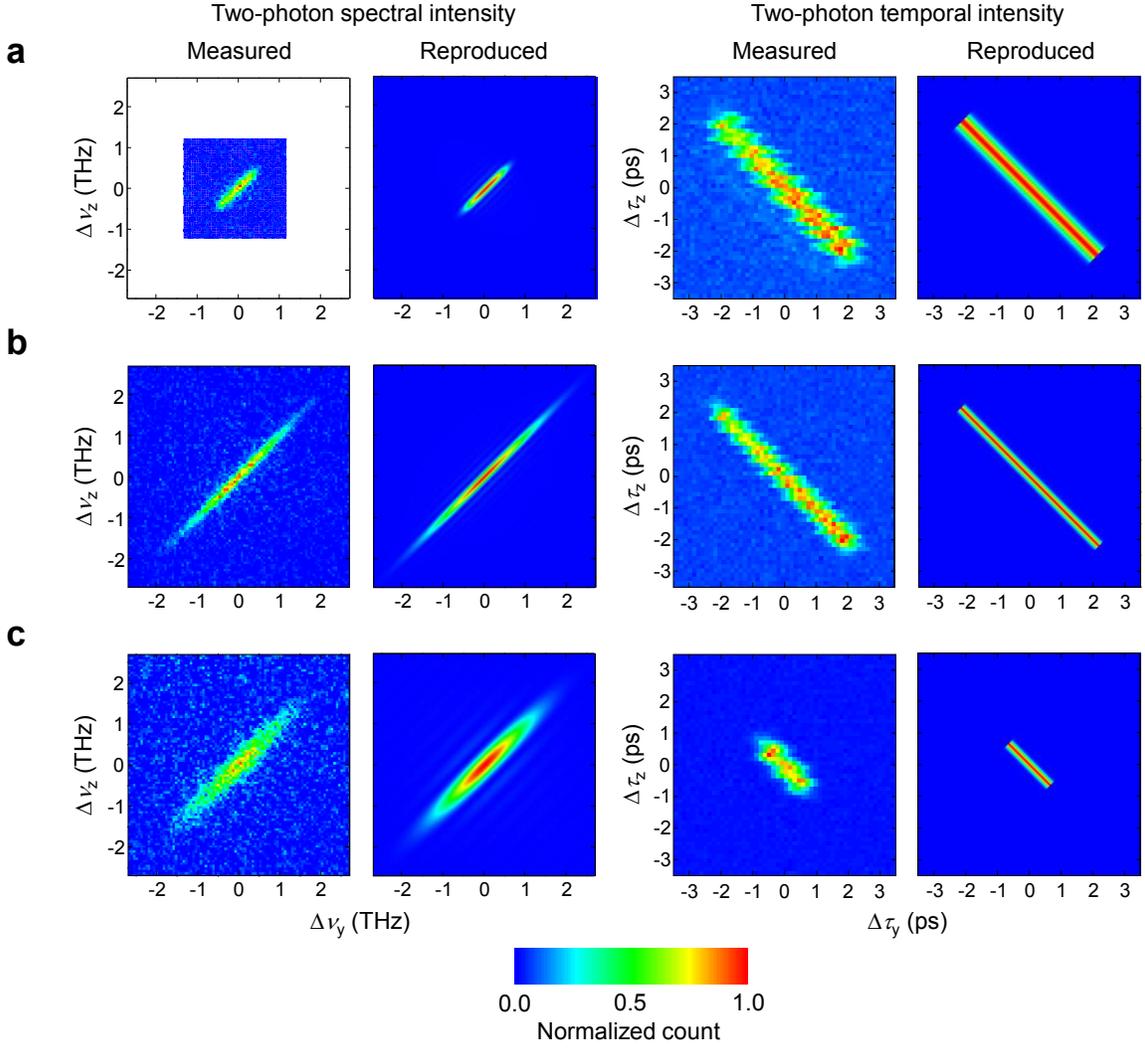}
\caption{ \textbf{Observation of time-frequency duality of biphotons.} The plots in the left two columns are TSIs, and those in the right two columns are the TTIs.
No measurements were carried out in the white area in \textbf{(a)}.
The biphotons were produced under different conditions.
\textbf{(a)} The pump had a near-rectangular profile with a bandwidth $\Delta \lambda_p $ of 2.8 nm and PPKTP crystal length $L$ of 30 mm;
\textbf{(b)} the pump had a Gaussian profile with a bandwidth $\Delta \lambda_p$ of 8.1 nm and $L$ of 30 mm;
\textbf{(c)} the pump also had a Gaussian profile with a bandwidth $\Delta \lambda_p$ of 8.1 nm and $L$ of 10 mm.
The temporal distributions varied by manipulating the spectral distributions.
}
\label{results}
\end{figure*}

Figure\,\ref{results} shows the experimentally measured TSIs (left) and TTIs (right) with a combination of the two different pump bandwidths (2.8 or 8.1 nm) and two different crystal lengths (10 or 30 mm).
We reproduced TSIs and TTIs deconvoluted from the original data, taking into account the resolutions of our two-photon detection systems (see Methods).
%
%
%
%
%
%
%
%
In Fig.\,\ref{results}, the horizontal (vertical) axis $\Delta \nu _y$ ($\Delta \nu _z$) is the frequency-shift from the center frequency of each distribution in the y- (z-)direction-polarized photons.
The zero-shifted frequencies are 189.4 THz, corresponding to a center wavelength of 1584 nm.
The specific features of the time-frequency duality of the biphotons can be clearly observed in Fig.\,\ref{results}; the TSI of the biphotons from the PPKTP crystal has a positively correlated distribution along the diagonal ($\Delta \nu_y = \Delta \nu _z$) direction, while the corresponding TTI has a negatively correlated distribution along the anti-diagonal ($\Delta \nu_y = -\Delta \nu _z$) direction.

When the pump bandwidth was increased from 2.8 nm in Fig.\,\ref{results}a to 8.1 nm in Fig.\,\ref{results}b for the fixed crystal length of 30 mm, the TSI became slightly broader along the diagonal direction, while the width was unchanged along the anti-diagonal direction.
In the same manner, when the crystal length $L$ was shortened from 30 mm in Fig.\,\ref{results}b to 10 mm in Fig.\,\ref{results}c, the TSI became broader along the anti-diagonal direction, while staying the same in the diagonal direction.
We can understand these phenomena from the following facts:
the biphoton spectral amplitude  $\tilde A^{(2)}(\Delta\nu _y, \Delta\nu _z)$ is the product of a phase-matching function $\Phi (\nu _-, L)$ and a pump spectral function $\alpha (\nu _+)$; $\tilde A^{(2)}(\nu _+ ,\nu _- ) = \Phi (\nu _-, L)\alpha (\nu _+)$, where $\nu_\pm = \frac{1}{\sqrt{2}}(\Delta \nu_y - \Delta \nu_z)$ and $L$ is the given PPKTP length.
Therefore the bandwidth of the TSI in the $\nu _+$ direction is determined by the spectral function of the pump laser, while that in the $\nu _-$ direction is determined by the phase-matching function depending on the crystal length $L$ (seeSupplementary Information).
Thus, in our case, the time-frequency duality of the biphoton can be expressed in the following form:
\begin{eqnarray}
\label{eq4}
A^{(2)}(\Delta \tau_y ,\Delta \tau_z) &=&  \frac{1}{\sqrt {2\pi }  }\int d\nu_+  \alpha (\nu _+)  e^{ - 2\pi i \nu_+ \tau_+} \nonumber \\
& & {} \times   \frac{1}{\sqrt {2\pi }  }\int d\nu_-  \Phi (\nu _-, L)  e^{ - 2 \pi i \nu_- \tau_-} \nonumber \\
&=&A^{(1)}_1(\tau_+)A^{(1)}_1(\tau_-).
\end{eqnarray}
where $\tau_\pm = \frac{1}{\sqrt{2}}(\Delta \tau_y \pm \Delta \tau_z)$.
This indicates that the 2DFT can be decomposed into the product of two 1DFTs: $A^{(1)}_1(\tau_+) =   \frac{1}{\sqrt {2\pi }  }\int d\nu_+  \alpha (\nu _+)  e^{ - 2\pi i \nu_+ \tau_+}$ and $A^{(1)}_2(\tau_-) = \frac{1}{\sqrt {2\pi }  }\int d\nu_-  \Phi (\nu _-, L)  e^{ - 2 \pi i \nu_- \tau_-}$.
Based on this, we confirmed the time-bandwidth product (TBP) between $\delta \tau_+$ and $\delta \nu_+$ or $\delta \tau_-$ and $\delta \nu_-$, where $\delta \tau_\pm$ ($\delta \nu_\pm$) is the full width at half maximum (FWHM) values along the $\tau_\pm$ ($\nu_\pm$) directions in the TTIs  (the TSIs).

\begin{figure}[tbhp]
\centering
\includegraphics[width= 0.45\textwidth]{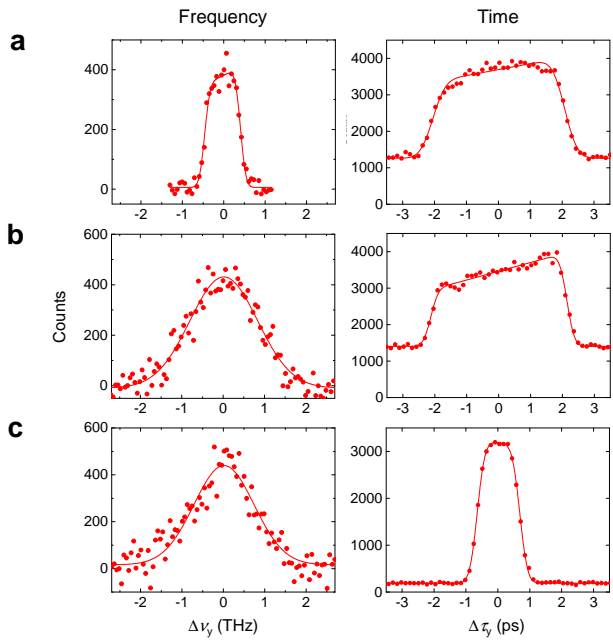}
\caption{ \textbf{Marginal distributions of two photon spectral and temporal intensity distributions.} Left graphs are the marginal distributions of TSIs, shown in Fig.\,\ref{results}, projecting onto the  $\Delta \nu_y$ axis.
Right graphs are those of TTIs, projecting onto the $\Delta \tau_y$ axis.
The red lines represent a least-square fit to the data.
Back ground counts  exist  in the TTI measurements due to the unwanted photon pair generation in the upconversion crystal.
No relations are observed for corresponding distributions between the time and frequency domains.
}
\label{marginal}
\end{figure}

Using the reproduced data in Fig.\,\ref{results}, we estimated $\delta \nu_\pm$ and  $\delta \tau_\pm$ for all the cases in Fig.\,\ref{results} and calculated TBPs.
It is worth evaluating the TBP for the marginal distributions, as shown in Fig.\,\ref{marginal}, because they can be observed by classical spectroscopic measurements.
Here, $\delta \nu_y (\delta \tau_y)$ is the FWHM of the marginal distribution of the TSI (TTI) after the deconvolution.
All TBP values are summarized in Table \,\ref{table1}.

It is obvious that the TBP values between $\delta \tau_\pm$ and $\delta \nu_\pm$ are less than 1, meaning that the generated biphoton wave packets satisfy the nearly Fourier transform limited condition.
The values of $\delta \nu_+ \times \delta \tau_+ $ are almost determined by the pump pulse spectra $\alpha (\nu _+)$, and would have the value of approximately 0.44 when assuming a Gaussian.
On the other hand, $\delta \nu_- \times \delta \tau_- $ is associated with the phase-matching function $\Phi (\nu _-, L)$, and the values would be approximately 0.87 due to the rectangular shape of the PPKTP crystals along the propagation direction of the pump pulse.
However, the value of $\delta \nu_y \times \delta \tau_y $ in Table \,\ref{table1} is clearly larger than 1.
This implies that the constituent photons of the biphoton wave packet no-longer satisfy the Fourier transform limited conditions (i.e., the coherence of  single-photon wave packets are degraded).
The disappearance of the single-photon coherence is the inherent characteristics in a quantum-mechanical biparticle system and is well-known as the degradation of purity in a qubit system \cite{Nielsen2000}.

\textbf{\emph{Discussion}}
We now consider the difference of time-frequency duality between classical and quantum regimes.
It is known that the Hamiltonian for a classical electromagnetic wave can be written as the sum of Hamiltonians for independent harmonic oscillators.
Thus, in terms of the photon picture, we can interpret classical optical phenomena as the collective behavior of uncorrelated photons, which results in a single photon (1D) treatment.
Under the 1D treatment, we understand the electric field amplitude, or one-photon probability amplitude, between time and frequency domains are conjugate physical quantities.
In contrast, the limitation of the 1D treatment is evident from the results presented in Figs.\,\ref{results} and\,\ref{marginal}.
In practice, the TBP of $\delta \nu_y \times \delta \tau_y $ implies the single-photon spectrum associated with the marginal distribution in the frequency domain is no longer conjugate with the single-photon temporal shape.
In our case, it is obvious that the marginal distributions in the frequency domain is mostly determined by the distribution along the diagonal direction in the TSIs, whose Fourier conjugate pair is the distribution along the diagonal direction in the time domain.
In other words, the single-photon spectral distribution strongly affects the temporal quantum correlation of the biphoton.
In the same manner, the frequency quantum correlation leads to the single-photon temporal distribution.
Through these considerations, we can understand classical optical theory only does not cause inconsistency for the collective behavior of uncorrelated photon such as laser pulses.
Higher-order treatments incorporates the quantum mechanics, such as a quantum many-body system \cite{Schweigler2017}, is essentially required for general understanding the nature of light.

\begin{table}[tbhp]
\centering
\begin{tabular}{|c|c|c|c|c|c|}
\hline
  & $\Delta \lambda _p$         & $L$                & $\delta \tau_+\times \delta \nu_+$     & $\delta \tau_-\times \delta \nu_-$  & $\delta \tau_y\times \delta \nu_y$\\
\hline
\textbf{a} & 2.8 nm       &30 mm   & 0.46   & 0.77   & 3.4\\
\hline
\textbf{b} & 8.1 nm       & 30 mm    & 0.49   & 0.85   & 8.2\\
\hline
\textbf{c} & 8.1 nm      & 10 mm      & 0.41   & 0.59   & 2.2\\
\hline
\end{tabular}
\caption{\label{table1}
\textbf{Time-bandwidth products of the biphoton wave packets.} Summary of  the time-bandwidth products in three experimental conditions in Fig.\,\ref{results}. All the values are less than 1 in the two-photon distributions, implying biphotons nearly satisfy the Fourier-limited conditions. On the other hand, time-bandwidth product values in the marginal distributions clearly donot meet the Fourier-limited condition.}
\end{table}

As a many-body quantum system, our study opens up possible new directions for optical science and technologies, taking into account the quantum correlation, which classical optics treats as the collective motion of uncorrelated photons.
Specifically, time-frequency duality with the higher-dimensional treatments has great potential to manage a multiple-photon wave packet in a higher dimensional time-frequency space.
Here, we only manipulated the real part of the two-photon spectral amplitude (i.e., the TSI). By modulating an imaginary two-photon spectral amplitude, i.e., the phase between two-photon spectral modes, in addition to the real part, however, we could easily extent our work to a new quantum optical synthesis technology based on a 2D Fourier optical treatment.
This may lead to a deeper understanding of the essential nature of light and to the development of future photonics at the single-photon level in the time-frequency domain.

\textbf{\emph{Conclusion}}
In summary, we have demonstrated the correlation inversion between time and frequency domains for frequency-positive-correlated biphotons, by experimentally measuring two-photon distributions in both domains. It was discovered that the generated biphotons satisfy the Fourier limited condition quantum mechanically, but not classically. Our study opens up possible new directions for optical science and technologies a many-body and high-dimensional frequency space.

\textbf{\emph{Acknowledgements}}
We thank Y. Guo for the helpful discussions. R.J. is supported by a fund from the Educational Department of Hubei Province, China (Grant No. D20161504), and by National Natural Science Foundations of China (Grant No.11704290). R.S. acknowledges support from the Research Foundation for Opto-Science and Technology, Hamamatsu, Japan, and support from Matsuo Foundation, Tokyo, Japan.

\onecolumngrid
\clearpage

\textbf{\emph{Methods}}

\begin{figure*}[tbhp]
\centering
\includegraphics[width= 0.95\textwidth]{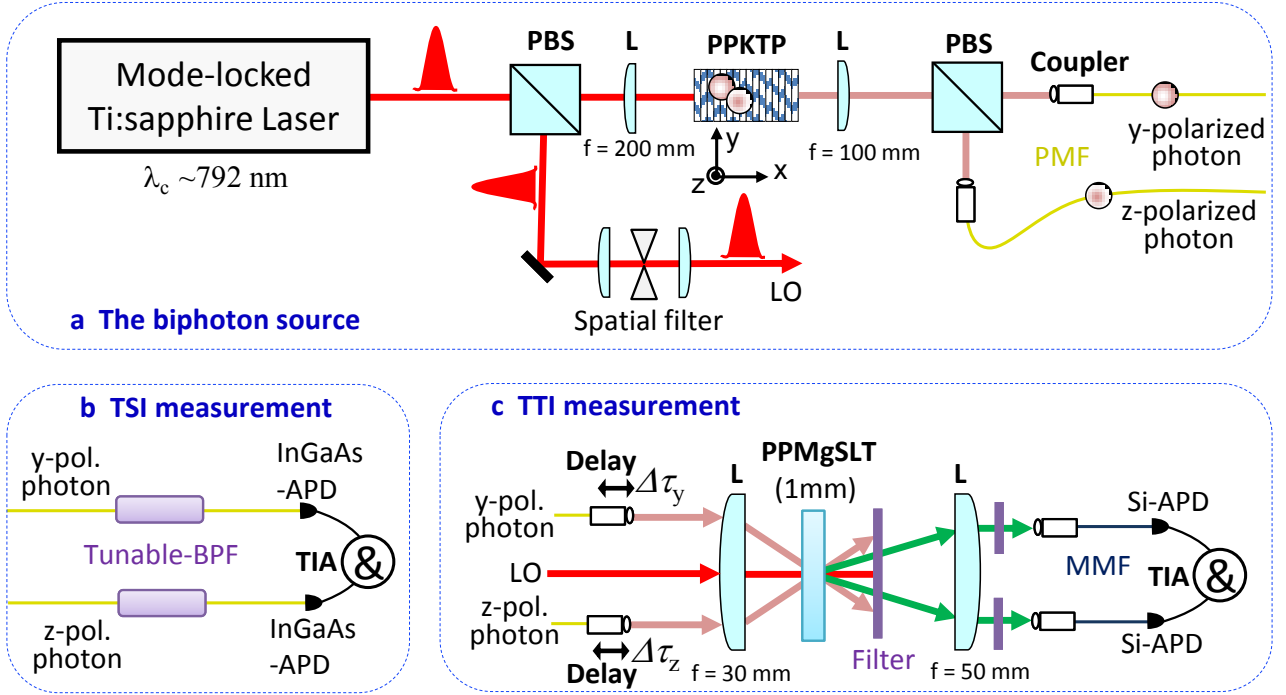}
\caption{ \textbf{ Experimental setup}. \textbf{(a)} shows how biphotons were generated; \textbf{(b)}  measurement system for two-photon spectral intensity (TSI); \textbf{(c)} measurement system fortwo-photon temporal intensity (TTI). L, lens; PBS, polarizing beam splitter; PMF, polarization maintaining fiber; BPF, bandpass filters;  APD, avalanche photodiode; LO, local oscillator; TIA, time-interval analyzer; \&, coincidence counter.
}
\label{setup}
\end{figure*}

\noindent
\textbf{Schematic illustration of the whole  experimental setup}
The experiment setup is shown in Fig.\,\ref{setup}. We used femtosecond laser pulses with a repetition rate of 76 MHz from a mode-locked titanium sapphire laser, operating at the center wavelength of 792 nm with a bandwidth of 8.1 nm. These were divided into two paths by a polarization beam splitter (PBS) in order to generate and detect biphotons.
For the biphoton generation, pulses were sent to a periodically poled  potassium titanyl phosphate (PPKTP) crystal with a poling period of 46.1 $\mu$m for type-II group-velocity-matched spontaneous parametric down-conversion. 
To manipulate the two-photon spectral distribution, we also prepared the pump pulse with a bandwidth of 2.8 nm (by inserting a bandpass filter), and crystals of 30 and 10 mm in length.
Thanks to the group velocity matching condition with the femtosecond laser pulse pumping, we could generate biphotons with positive frequency correlations at the center wavelength of 1584 nm.
Since biphotons generated via the type-II phase-matching condition have orthogonal polarizations, the polarization of the constituent photons were aligned along either the crystallographic y- or z-axis.
The constituent photons were separated by a PBS and coupled into  two polarization-maintaining fibers (PMF).
The other path of the pump laser pulse was spatially filtered, and then used as a local oscillator (LO) for TTI measurement.
For TSI measurement \cite{Shimizu2009,Jin2013OE}
, each constituent photon of the biphoton was sent to tunable bandpass filters (BPFs), which had a  Gaussian-shaped filter function  with a fixed FWHM of 0.56 nm and a tunable central wavelength from 1560-1620 nm, followed by a two-photon detector consisting of two fiber-coupled single-photon detectors and a time interval analyzer (TIA).
The TSI was measured by scanning the central wavelength of the two BPFs with a step length of 0.5 nm,  and recording the coincidence counts for each point.
For TTI measurements, we utilized a time-resolved upconversion detection system with spatial multiplexing \cite{Kuzucu2008OL, Kuzucu2008PRL}.
A periodically poled MgO-doped stoichiometric lithium tantalate (PPMgSLT) crystal (poling period of 8.5 $\mu$m; length 1 mm) under the type-0 phase-matching condition  was used for  noncollinear sum-frequency generation (1584 nm + 792 nm $\to$ 528 nm).
After filtering the backgrounds, the up-converted photons were detected by two Si avalanche photodiodes (Si-APD),  which were connected to a TIA for  coincidence counting.
The TTI was measured by scanning the temporal delay between the constituent photons and the LO pulse with a step length of 0.13 ps, and we recorded coincidence events.
%
%
%
%
The temporal width of the LO pulse was estimated to be approximately 0.12 ps from the bandwidth of the pump pulse. The group-velocity difference between the LO pulse and the photons passing through the PPMgSLT was calculated to be 0.22 ps. We therefore estimated the temporal resolution for a one-photon detection with Si-APD to be $0.12 + 0.22 = 0.34$ ps, and for a two-photon detection, $0.34 \times \sqrt{2} \approx 0.48$ ps.

\begin{figure}[tbhp]
\centering
\includegraphics[width= 0.45\textwidth]{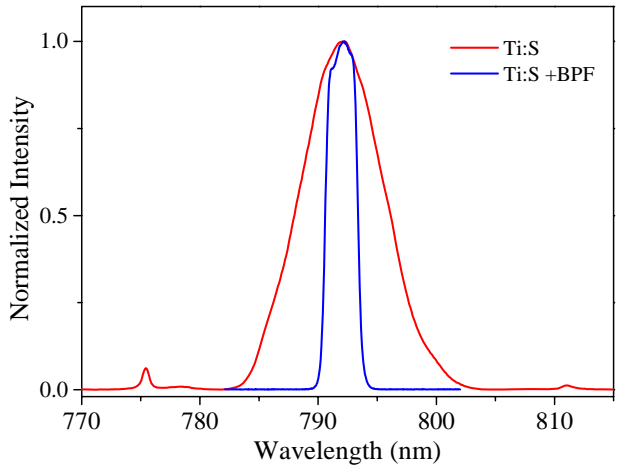}
\caption{ \textbf{Pump spectra:} \textbf{(a)} red¡ªthe original spectrum from the  mode-locked titanium sapphire laser;
\textbf{(b)} blue¡ªthe spectrum after the insertion of a BPF.
}
\label{pump}
\end{figure}

\noindent
\textbf{Pump pulse spectra for 2.8-nm and 8.1-nm bandwidths}
In our experiment, the femtosecond laser pulse had a Gaussian distribution with a center wavelength of 792 nm and bandwidth of 8.1 nm, as shown in Fig.\,\ref{pump} (red curve). This spectrum was used for the TSIs and TTIs in Figs.\,\ref{results}b and c in the main text. After the insertion of a bandpass filter, the bandwidth became 2.8 nm, while the center wavelength was unchanged, as shown in Fig.\,\ref{pump} (blue curve). This spectrum was used for the results in Fig.\,\ref{results}a in the main text.
%
%
%
%

\begin{figure*}[tbhp]
\centering
\includegraphics[width= 0.9\textwidth]{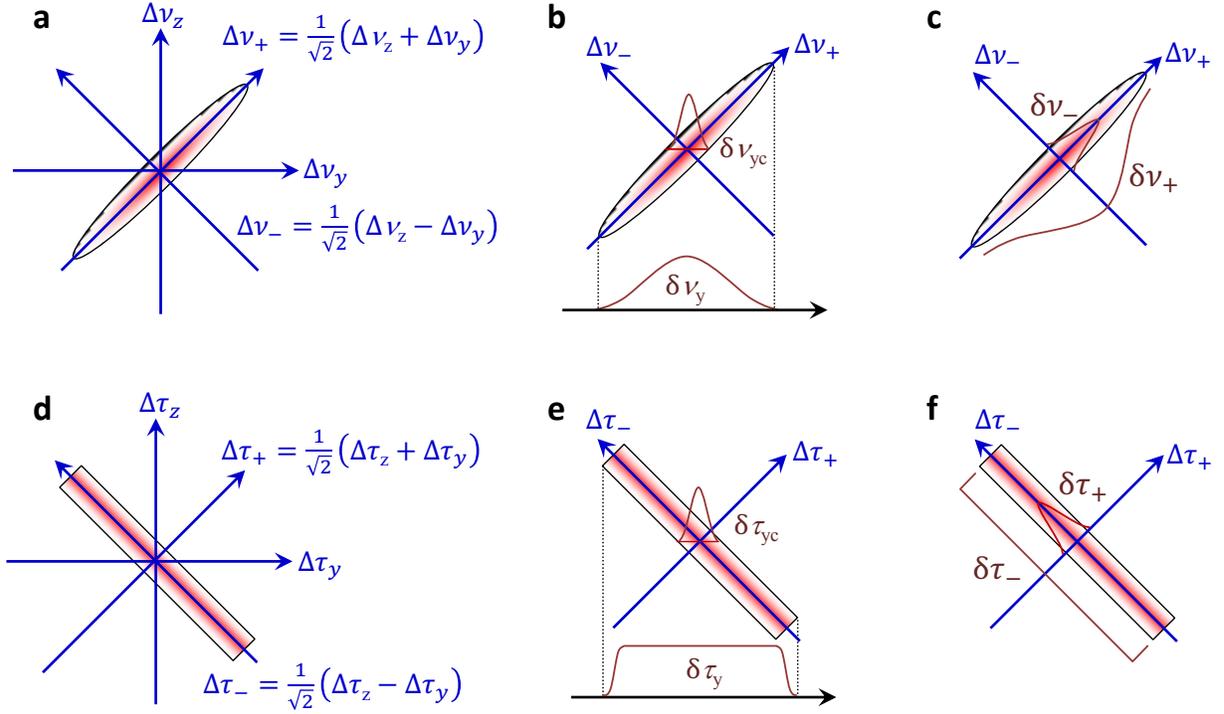}
\caption{ \textbf{Coordinates and parameters}. \textbf{(a)} The coordinate for the frequency domain; \textbf{(b)} $\delta \nu _{yc}$ is the FWHM of the distribution along the $\Delta \nu_y$ direction; $\delta \nu_{y}$ is the FWHM of the projected distribution onto the $\Delta \nu_y$ direction; \textbf{(c)} $\delta \nu _+$ is the FWHM of the projected distribution onto the $\nu_+$ direction; $\delta \nu_-$ is the FWHM of the projected distribution onto the $\nu_-$ direction;
\textbf{(d)} the coordinate for the time domain; \textbf{(e)} $\delta \tau _{yc}$ is the FWHM of the distribution along the $\Delta \tau_y$ direction; $\delta \tau_{y}$ is the FWHM of the projected distribution onto the $\Delta \tau_y$ direction;\textbf{ (f)} $\delta \tau _+$ is the FWHM of the projected distribution onto the $\tau_+$ direction; $\delta \tau_-$ is the FWHM of the projected distribution onto the $\tau_-$ direction.
}
\label{tband}
\end{figure*}

\noindent
\textbf{Coordinates and parameters}
We summarize the coordinates and parameters used in this paper, as shown in Figs.\,\ref{tband}(a-f).
$\Delta \nu_y$,  $\Delta \nu_z$, $\nu_+$ and $\nu_-$ are the horizontal, vertical, diagonal and anti-diagonal axes respectively in the frequency domain; $\Delta \tau_y$,  $\Delta \tau_z$, $\tau_+$ and $\tau_-$ are the horizontal, vertical, diagonal and anti-diagonal axes,  respectively, in the time domain; $\delta \nu_{yc} $ is the FWHM for the distribution long cross section of $\Delta \nu_z = 0$; meanwhile, $\delta \nu_y $, $\delta \nu_+ $ and $\delta \nu_- $  are the FWHM for the marginal distribution by projecting the data onto the axes of $\Delta \nu_y $, $\nu_+ $ and $\nu_- $ respectively.
Finally, $\delta \tau_{yc} $ is the FWHM for the distribution along the cross section of $\Delta \tau_z = 0$; and $\delta \tau_y $, $\delta \tau_+ $ and $\delta \tau_- $  are the FWHM for the marginal distribution by projecting the data onto the axes of $\Delta \tau_y $, $\tau_+ $ and $\tau_- $ respectively.
%

\begin{figure*}[tbhp]
\centering
\includegraphics[width= 0.9\textwidth]{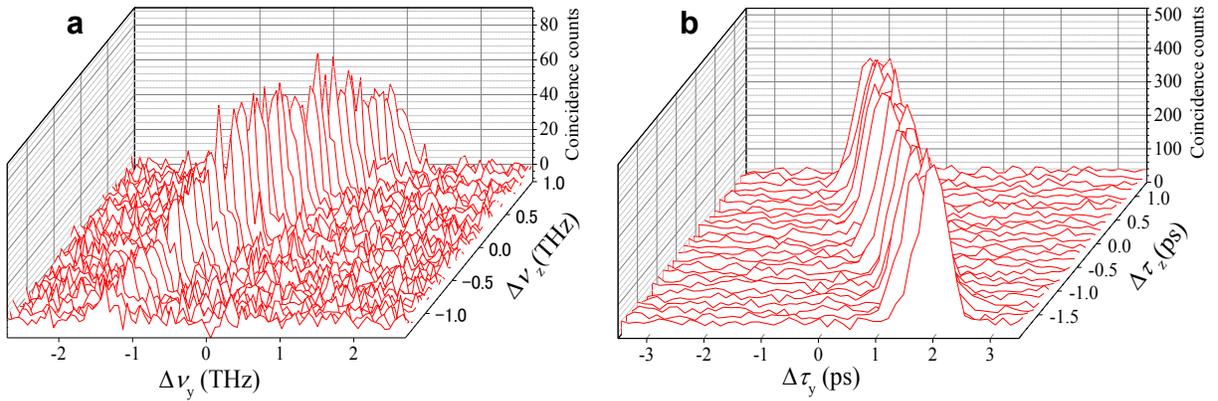}
\caption{ \textbf{Detailed analysis of the TSI and TTI}. \textbf{(a)} The waterfall plots of the TSI;  and \textbf{(b)} TTI in Fig.\,\ref{results}b.
}
\label{3Dtsitti}
\end{figure*}

\noindent
\textbf{Reproduction of the TSIs and TTIs}
We take Fig.\,\ref{results}b in the main text as an example to show the details of the experimental data processing method.
The image on the left in Fig.\,\ref{results}b is the density plot of the TSI, which was scanned along the $\Delta\nu_y$ direction at each $\Delta\nu_z$ value.
Figure\,\ref{3Dtsitti}a is the waterfall plot of Fig.\,\ref{results}b.
We fitted the distributions along the $\Delta\nu_z$ direction using Gaussian functions, and obtained an averaged FWHM value of $\delta \nu_{yc} = 0.21$ THz.
Next, we projected the TSI data onto the axis of $\Delta\nu_y$, and obtained the marginal distribution, shown in Fig.\,\ref{marginal}b.
The FWHM value of $\delta \nu_{y} = 1.9 $ THz was estimated with a Gaussian function.
%
%
Similar results can also be obtained for the temporal data.
The density plot of the TTI in Fig.\,\ref{results}b was scanned by 76$\times$76 steps.
Figure\,\ref{3Dtsitti}b is the waterfall plot of the TTI.
We fitted the distributions along the $\Delta\tau_z$ direction using Gaussian functions, and obtained a FWHM of $\delta \tau_{yc} = 0.26$ ps.
From the projection of the TTI onto the  $\Delta\tau_y$ axis, we obtained the marginal distribution, as shown in Fig.\,\ref{marginal}b.
With a rectangular-like profile, we estimated a FWHM value of $\delta \tau_{y} = 4.3 $ ps.
The other four data measurements in Figs.\,\ref{results}a and c were processed similarly.
Using all the FWHM values, we reproduced the TSIs and TTIs in Fig.\,\ref{results}.
We summarize the parameters for TSI and TTI in Table \,\ref{table2}.

\begin{table*}[tbhp]
\centering
\begin{tabular}{|c|c|c|c|c|c|c|c|c|c|c|}
\hline
\multicolumn{3}{|c|}{ }&\multicolumn{4}{c}{Measured}&\multicolumn{4}{|c|}{Reproduced} \\
\hline
\multicolumn{3}{|c|}{Width}&$\delta \nu_y$&$\delta \nu_{yc}$&$\delta \tau_y$&$\delta \tau_{yc}$ &$\delta \nu_+$&$\delta \nu_-$&$\delta \tau_+$&$\delta \tau_-$ \\
\hline
\multicolumn{3}{|c|}{Unit}&(THz)&(THz)&(ps)&(ps) &(THz)&(THz)&(ps)&(ps) \\
\hline
\textbf{a} & $\Delta \lambda _p$=2.8 nm  &$L$=30 mm   &0.82 &0.19 &4.2 &0.54 &1.2 &0.13 &0.38 &5.9\\
\hline
\textbf{b} & $\Delta \lambda _p$=8.1 nm  & $L$=30 mm  &1.9  &0.21 &4.3 &0.26 &2.7 &0.14 &0.18 &6.1\\
\hline
\textbf{c} & $\Delta \lambda _p$=8.1 nm  & $L$=10 mm  &1.7  &0.49 &1.3 &0.25 &2.3 &0.33 &0.18 &1.8 \\
\hline
\end{tabular}
\caption{\label{table2}
\textbf{The parameters for TSI and TTI}. Summary of  the  bandwidths for measured and reproduced TSI/TTI.}
\end{table*}

\noindent
\textbf{Measurement of TSI}
The TSI was measured  using two center-wavelength-tunable BPFs, which had a Gaussian-shaped filter function  with an FWHM of 0.56 nm and a tunable central wavelength from 1560-1620 nm \cite{Shimizu2009, Jin2013OE, Bisht2015}.
The two single photon detectors used in this measurement were two InGaAs avalanche photodiode (APD) detectors (ID210, idQuantique), which had  a quantum efficiency of around 20\%, with a dark count of around 2 kHz.
To measure the TSI of the photon pairs, we scanned the central wavelength of the two BPFs, and recorded the coincidence counts. The two BPFs were moved  0.1-nm per step and 60$\times$60 steps in all. The coincidence counts were accumulated for 5 s for each point.

\clearpage

\subsection*{Supplementary Information}

\begin{figure*}[tbh]
\centering
\includegraphics[width= 0.95\textwidth]{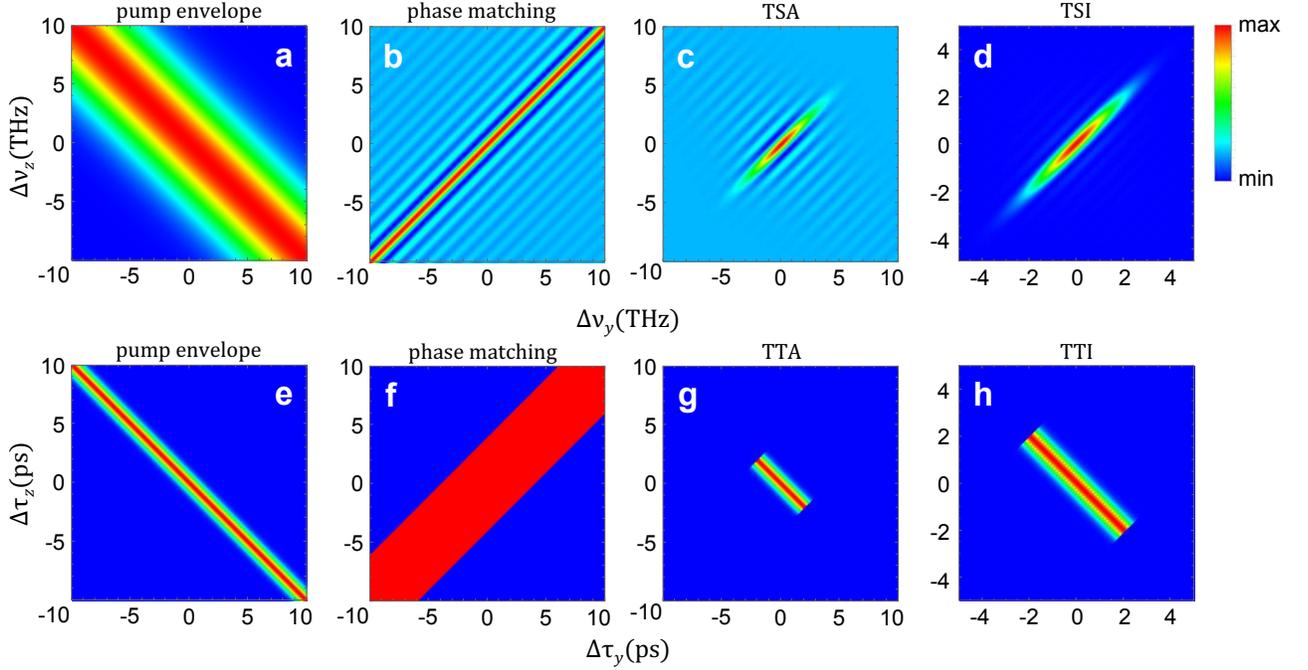}
\caption{ \textbf{The mathematical model}. The first row is for spectral distribution: \textbf{(a)}  pump envelope function $\exp[ - a(\nu _s  + \nu _i )^2 ]$; \textbf{(b)} phase matching function ${\rm sinc}[b(\nu _s  - \nu _i )]$; \textbf{(c)}  two-photon spectral amplitude (TSA)  $f(\nu_s, \nu_i)$; \textbf{(d)}  two-photon spectral intensity (TSI) $|f(\nu_s, \nu_i)|^2$.  The second row  shows the corresponding temporal distributions: \textbf{(e)}  temporal pump envelope function $\exp[ - \frac{(t_s  + t_i )^2}{16a}]$; \textbf{(f)}  temporal phase matching function ${ \rm rect}(\frac{t_s  - t_i}{2b})$; \textbf{(g)}  two-photon temporal amplitude (TTA) $F(t_s, t_i)$; \textbf{(h)}  two-photon temporal intensity (TTI) $|F(t_s, t_i)|^2$. The parameters are set at $a=0.04$ and $b=4$ for all the figures.
}
\label{MathModel}
\end{figure*}

\textbf{S1:  Theoretical model of TSI and TTI}
For a  deepper understanding of this experiment, we constructed a simple but effective mathematical model for the biphotons from the PPKTP crystal.
The two-photon spectral amplitude (TSA), $f(\nu_s, \nu_i)$,  is the product of the pump-envelope function and phase-matching function.
For simplicity in the calculation, we define $\nu_s \equiv \Delta \nu_y$ and $\nu_i \equiv \Delta \nu_y$ as  the shifted-frequencies of the signal and idler photons in the main text.
Without loss of generality, we assume the pump-envelope function as $\alpha(\nu_s + \nu_i)= \exp [ - a(\nu _s  + \nu _i )^2 ]$, where $a$ is determined by the spectral width of the pump laser. $\alpha(\nu_s + \nu_i)$ has a distribution 135-degrees to the horizontal axis (i.e., it is distributed along the anti-diagonal direction, as shown in Fig.\,\ref{MathModel}a).
The phase-matching function is $\Phi(\omega_s, \omega_i)={\rm sinc}[b(\nu _s  - \nu _i )]$, where $b$ is determined by the length of the PPKTP crystal.
The PPKTP crystal satisfies the GVM condition at telecom wavelength \cite{Jin2013OE, Konig2004, Evans2010, Gerrits2011, Eckstein2011}
, therefore, its phase-matching function has a distribution 45-degrees to the horizontal axis (i.e., distributed along the diagonal direction, as shown in Fig.\,\ref{MathModel}b).
The TSA can be written as $f(\nu_s, \nu_i)= \exp [ - a(\nu _s  + \nu _i )^2 ] {\rm sinc}[b(\nu _s  - \nu _i )]$, which is the product of the pump-envelope function and phase-matching function. In this experiment, the  pump envelope is wider than the width of the phase-matching function; therefore,  their product also distributes along the diagonal direction. That is, the signal and idler are positively correlated, as shown in Fig.\,\ref{MathModel}c.
The two-photon spectral intensity is $|f(\nu_s, \nu_i)|^2$, as shown in Fig.\,\ref{MathModel}d.
%
%
%
%
After the long calculation  described in \textbf{S2}, the Fourier transform of $f(\nu_s, \nu_i)$ can be analytically obtained as
$F(t_s, t_i)= c_0 \exp[ - \frac{(t_s  + t_i )^2}{16a}] {\rm rect}(\frac{t_s  - t_i}{2b})$, where $c_0=\frac{2\pi ^3 \sqrt \pi  }{b\sqrt a }$.
Note that we define $t_s \equiv \Delta\tau_y$ and $t_i \equiv \Delta \tau_z$ as the shifted-time for the signal and idler photons in the main text.
Here, $\exp[-\frac{(t_s  + t_i )^2}{16a}]$ is the pump-envelope function in the time domain; ${\rm rect}(\frac{t_s  - t_i}{2b})$ is the phase-matching function in the time domain; $F(t_s, t_i)$ is the two-photon temporal amplitude (TSA) and   $|F(t_s, t_i)|^2$ is just the two-photon temporal intensity (TTI). The distribution of these functions is shown in Figs.\,\ref{MathModel}e - h.
It is interesting to note that, $\nu _s$ and $\nu _i$ can not be written in a separated form in $f(\nu_s, \nu_i)$. However, if we introduce new variables $\nu _\pm= \frac{1}{{\sqrt 2 }}(\nu_s \pm \nu_i)$, $\nu _+$ and $\nu _-$ can be separated in $f(\nu_+, \nu_-)$. The situation is similar for $\tau _s$ and $\tau _i$, as explained in Eq.\,(3) in the main text. Another interesting feature is that the pump-envelope  and  phase-matching functions are independent from each other in both the spectral and temporal domains.
The spectral and temporal phase-matching functions are varied simultaneously by changing the parameter a, while phase-matching functions are not affected. Similar phenomena can be observed for the spectral and temporal pump-envelope functions by changing the parameter b:  varying  the $sinc$ function in the spectral domain corresponds to varying the $rect$ function in the time domain, while the Gaussian-shaped spectral and temporal  pump-envelope functions are not affected.

This model is used for the simulations of Fig.\,\ref{results} in the main text. For  Fig.\,\ref{results}a, we changed the pump-envelope function from a Gaussian function to  a convolution between a Gaussian function and rectangular function, because the pump laser for Fig.\,\ref{results}a was filtered by a near-rectangular shaped BPF.

\noindent
\textbf{S2: Theoretical calculation of TTA from TSA}

For a pump-envelope function $\exp[ - a(\nu _s  + \nu _i )^2 ]$ and a phase-matching  function ${\rm sinc}[b(\nu _s  - \nu _i )]$,
the two-photon spectral amplitude (TSA) can be written as
\begin{equation}
\label{eqs3}
f(\nu _s, \nu _i)=\exp[ - a(\nu _s  + \nu _i )^2 ] {\rm sinc}[b(\nu _s  - \nu _i )]
\end{equation}
The two-photon temporal amplitude (TTA) is the Fourier transform ($\mathscr{F}$) of $f(\nu _s, \nu _i)$, and can be calculated as follows.
\begin{equation}
\label{eqs5}
\begin{array}{lll}
 \mathscr{F}\{ f(\nu _s ,\nu _i )\}  &=& \mathscr{F} \{ \exp [ - a(\nu _s  + \nu _2 )^2 ]{\rm sinc} [b(\nu _s  - \nu _2 )]\}  \\ \\
  &=& \mathscr{F} \{ \exp [ - a(\nu _s  + \nu _i )^2 ]\}  \otimes \mathscr{F} \{ {\rm sinc} [b(\nu _s  - \nu _i )]\}  \\ \\
  &=&[2\pi \sqrt {\frac{\pi }{a}} \exp ( - \frac{{t_s^2 }}{{4a}})\delta (t_s  - t_i )] \otimes [2\pi ^2 \frac{1}{b} {\rm rect} (\frac{{t_s }}{b})\delta (t_s  + t_i )] \\ \\
  &=&\int_{ - \infty }^\infty  {\int_{ - \infty }^\infty  {d\tau _s d\tau _i [} } 2\pi \sqrt {\frac{\pi }{a}} \exp ( - \frac{{\tau _s^2 }}{{4a}})\delta (\tau _s  - \tau _i )][2\pi ^2 \frac{1}{b} {\rm rect} (\frac{{t_s  - \tau _s }}{b})\delta (t_s  + t_i  - \tau _s  - \tau _i )] \\ \\
  &=&\int_{ - \infty }^\infty  {d\tau _s [} 2\pi \sqrt {\frac{\pi }{a}} \exp ( - \frac{{\tau _s^2 }}{{4a}})][2\pi ^2 \frac{1}{b} { \rm rect} (\frac{{t_s  - \tau _s }}{b})\delta (t_s  + t_i  - \tau _s  - \tau _s )] \\ \\
  &=&\int_{ - \infty }^\infty  {d\tau _s [} 2\pi \sqrt {\frac{\pi }{a}} \exp({ - \frac{{\tau _s^2 }}{{4a}}}) ][2\pi ^2 \frac{1}{b} {\rm rect} (\frac{{t_s  - \tau _s }}{b})\frac{1}{2}\delta (\tau _s  - \frac{{t_s  + t_i }}{2})] \\ \\
  &=& \frac{{2\pi ^3 }}{b}\sqrt {\frac{\pi }{a}} \exp [ - \frac{{(t_s  + t_i )^2 }}{{16a}}] {\rm rect} (\frac{{t_s  - t_i }}{{2b}}) \\ \\
  &=& c_0 \exp [ - \frac{{(t_s  + t_i )^2 }}{{16a}}] {\rm rect} (\frac{{t_s  - t_i }}{{2b}}), \\ \\
 \end{array}
 \end{equation}
where $\otimes$ is the convolution symbol and $c_0 =\frac{{2\pi ^3 \sqrt \pi  }}{{b\sqrt a }}$. In conclusion, the TTA ($F(t_s, t_i)$) can be calculated from the Fourier transform of TSA ($f(\nu _s, \nu _i)$) with the following form:
\begin{equation}
\label{eqs6}
F(t_s, t_i)=\mathscr{F} \{f(\nu _s, \nu _i)\}=c_0 \exp[ - \frac{{(t_s  + t_i )^2 }}{{16a}}]{\rm rect} (\frac {t_s  - t_i}{2b}).
 \end{equation}


\begin{thebibliography}{38}%
\makeatletter
\providecommand \@ifxundefined [1]{%
 \@ifx{#1\undefined}
}%
\providecommand \@ifnum [1]{%
 \ifnum #1\expandafter \@firstoftwo
 \else \expandafter \@secondoftwo
 \fi
}%
\providecommand \@ifx [1]{%
 \ifx #1\expandafter \@firstoftwo
 \else \expandafter \@secondoftwo
 \fi
}%
\providecommand \natexlab [1]{#1}%
\providecommand \enquote  [1]{``#1''}%
\providecommand \bibnamefont  [1]{#1}%
\providecommand \bibfnamefont [1]{#1}%
\providecommand \citenamefont [1]{#1}%
\providecommand \href@noop [0]{\@secondoftwo}%
\providecommand \href [0]{\begingroup \@sanitize@url \@href}%
\providecommand \@href[1]{\@@startlink{#1}\@@href}%
\providecommand \@@href[1]{\endgroup#1\@@endlink}%
\providecommand \@sanitize@url [0]{\catcode `\\12\catcode `\$12\catcode
  `\&12\catcode `\#12\catcode `\^12\catcode `\_12\catcode `\%12\relax}%
\providecommand \@@startlink[1]{}%
\providecommand \@@endlink[0]{}%
\providecommand \url  [0]{\begingroup\@sanitize@url \@url }%
\providecommand \@url [1]{\endgroup\@href {#1}{\urlprefix }}%
\providecommand \urlprefix  [0]{URL }%
\providecommand \Eprint [0]{\href }%
\providecommand \doibase [0]{http://dx.doi.org/}%
\providecommand \selectlanguage [0]{\@gobble}%
\providecommand \bibinfo  [0]{\@secondoftwo}%
\providecommand \bibfield  [0]{\@secondoftwo}%
\providecommand \translation [1]{[#1]}%
\providecommand \BibitemOpen [0]{}%
\providecommand \bibitemStop [0]{}%
\providecommand \bibitemNoStop [0]{.\EOS\space}%
\providecommand \EOS [0]{\spacefactor3000\relax}%
\providecommand \BibitemShut  [1]{\csname bibitem#1\endcsname}%
\let\auto@bib@innerbib\@empty
\bibitem [{\citenamefont {Stark}(1982)}]{Stark1982}%
  \BibitemOpen
  \bibfield  {author} {\bibinfo {author} {\bibfnamefont {Henry}\ \bibnamefont
  {Stark}},\ }\href@noop {} {\emph {\bibinfo {title} {Applications of optical
  Fourier transforms}}}\ (\bibinfo  {publisher} {Academic Press},\ \bibinfo
  {year} {1982})\BibitemShut {NoStop}%
\bibitem [{\citenamefont {Goodman}(2017)}]{Goodman2017}%
  \BibitemOpen
  \bibfield  {author} {\bibinfo {author} {\bibfnamefont {Joseph~W.}\
  \bibnamefont {Goodman}},\ }\href@noop {} {\emph {\bibinfo {title}
  {Introduction to Fourier Optics}}},\ \bibinfo {edition} {4th}\ ed.\ (\bibinfo
   {publisher} {Freeman, W. H.},\ \bibinfo {year} {2017})\BibitemShut {NoStop}%
\bibitem [{\citenamefont {Zhang}\ \emph {et~al.}(2014)\citenamefont {Zhang},
  \citenamefont {Wei}, \citenamefont {Marhic},\ and\ \citenamefont
  {Wong}}]{Zhang2014}%
  \BibitemOpen
  \bibfield  {author} {\bibinfo {author} {\bibfnamefont {Chi}\ \bibnamefont
  {Zhang}}, \bibinfo {author} {\bibfnamefont {Xiaoming}\ \bibnamefont {Wei}},
  \bibinfo {author} {\bibfnamefont {Michel~E.}\ \bibnamefont {Marhic}}, \ and\
  \bibinfo {author} {\bibfnamefont {Kenneth K.~Y.}\ \bibnamefont {Wong}},\
  }\bibfield  {title} {\enquote {\bibinfo {title} {Ultrafast and versatile
  spectroscopy by temporal fourier transform},}\ }\href@noop {} {\bibfield
  {journal} {\bibinfo  {journal} {Sci. Rep.}\ }\textbf {\bibinfo {volume}
  {4}},\ \bibinfo {pages} {5351} (\bibinfo {year} {2014})}\BibitemShut
  {NoStop}%
\bibitem [{\citenamefont {Smith}(2011)}]{Smith2011}%
  \BibitemOpen
  \bibfield  {author} {\bibinfo {author} {\bibfnamefont {Brian~C.}\
  \bibnamefont {Smith}},\ }\href@noop {} {\emph {\bibinfo {title} {Fundamentals
  of Fourier Transform Infrared Spectroscopy}}},\ \bibinfo {edition} {2nd}\
  ed.\ (\bibinfo  {publisher} {CRC Press},\ \bibinfo {year} {2011})\BibitemShut
  {NoStop}%
\bibitem [{\citenamefont {Bell}(1972)}]{Robert1972}%
  \BibitemOpen
  \bibfield  {author} {\bibinfo {author} {\bibfnamefont {Robert}\ \bibnamefont
  {Bell}},\ }\href@noop {} {\emph {\bibinfo {title} {Introductory Fourier
  Transform Spectroscopy}}},\ \bibinfo {edition} {1st}\ ed.\ (\bibinfo
  {publisher} {Academic Press},\ \bibinfo {year} {1972})\BibitemShut {NoStop}%
\bibitem [{\citenamefont {Sirohi}(2017)}]{Sirohi2017}%
  \BibitemOpen
  \bibfield  {author} {\bibinfo {author} {\bibfnamefont {Rajpal}\ \bibnamefont
  {Sirohi}},\ }\href@noop {} {\emph {\bibinfo {title} {Optical Methods of
  Measurement: Whole field Techniques}}},\ \bibinfo {edition} {2nd}\ ed.\
  (\bibinfo  {publisher} {CRC Press},\ \bibinfo {year} {2017})\BibitemShut
  {NoStop}%
\bibitem [{\citenamefont {Griffiths}\ and\ \citenamefont
  {De~Haseth}(2007)}]{Griffiths2007}%
  \BibitemOpen
  \bibfield  {author} {\bibinfo {author} {\bibfnamefont {Peter~R.}\
  \bibnamefont {Griffiths}}\ and\ \bibinfo {author} {\bibfnamefont {James~A.}\
  \bibnamefont {De~Haseth}},\ }\href@noop {} {\emph {\bibinfo {title} {Fourier
  Transform Infrared Spectrometry}}},\ \bibinfo {edition} {2nd}\ ed.\ (\bibinfo
   {publisher} {Wiley-Interscience},\ \bibinfo {year} {2007})\BibitemShut
  {NoStop}%
\bibitem [{\citenamefont {VanderLugt}(2005)}]{VanderLugt2005}%
  \BibitemOpen
  \bibfield  {author} {\bibinfo {author} {\bibfnamefont {Anthony}\ \bibnamefont
  {VanderLugt}},\ }\href@noop {} {\emph {\bibinfo {title} {Optical Signal
  Processing}}}\ (\bibinfo  {publisher} {Wiley-Interscience},\ \bibinfo {year}
  {2005})\BibitemShut {NoStop}%
\bibitem [{\citenamefont {Rhodes}(2009)}]{Rhodes2009}%
  \BibitemOpen
  \bibfield  {author} {\bibinfo {author} {\bibfnamefont {W.T.}\ \bibnamefont
  {Rhodes}},\ }\href@noop {} {\emph {\bibinfo {title} {Fourier Optics and
  Optical Signal Processing}}}\ (\bibinfo  {publisher} {Wiley-Blackwell},\
  \bibinfo {year} {2009})\BibitemShut {NoStop}%
\bibitem [{\citenamefont {Chan}\ \emph {et~al.}(2011)\citenamefont {Chan},
  \citenamefont {Hsieh}, \citenamefont {Liang}, \citenamefont {Kung},
  \citenamefont {Lee}, \citenamefont {Lai}, \citenamefont {Pan},\ and\
  \citenamefont {Peng}}]{Chan2011}%
  \BibitemOpen
  \bibfield  {author} {\bibinfo {author} {\bibfnamefont {Han-Sung}\
  \bibnamefont {Chan}}, \bibinfo {author} {\bibfnamefont {Zhi-Ming}\
  \bibnamefont {Hsieh}}, \bibinfo {author} {\bibfnamefont {Wei-Hong}\
  \bibnamefont {Liang}}, \bibinfo {author} {\bibfnamefont {A.~H.}\ \bibnamefont
  {Kung}}, \bibinfo {author} {\bibfnamefont {Chao-Kuei}\ \bibnamefont {Lee}},
  \bibinfo {author} {\bibfnamefont {Chien-Jen}\ \bibnamefont {Lai}}, \bibinfo
  {author} {\bibfnamefont {Ru-Pin}\ \bibnamefont {Pan}}, \ and\ \bibinfo
  {author} {\bibfnamefont {Lung-Han}\ \bibnamefont {Peng}},\ }\bibfield
  {title} {\enquote {\bibinfo {title} {Synthesis and measurement of ultrafast
  waveforms from five discrete optical harmonics},}\ }\href@noop {} {\bibfield
  {journal} {\bibinfo  {journal} {Science}\ }\textbf {\bibinfo {volume}
  {331}},\ \bibinfo {pages} {1165} (\bibinfo {year} {2011})}\BibitemShut
  {NoStop}%
\bibitem [{\citenamefont {Suhaimi}\ \emph {et~al.}(2015)\citenamefont
  {Suhaimi}, \citenamefont {Ohae}, \citenamefont {Gavara}, \citenamefont
  {Nakagawa}, \citenamefont {Hong},\ and\ \citenamefont
  {Katsuragawa}}]{Suhaimi2015}%
  \BibitemOpen
  \bibfield  {author} {\bibinfo {author} {\bibfnamefont {Nurul~Sheeda}\
  \bibnamefont {Suhaimi}}, \bibinfo {author} {\bibfnamefont {Chiaki}\
  \bibnamefont {Ohae}}, \bibinfo {author} {\bibfnamefont {Trivikramarao}\
  \bibnamefont {Gavara}}, \bibinfo {author} {\bibfnamefont {Kenichi}\
  \bibnamefont {Nakagawa}}, \bibinfo {author} {\bibfnamefont {Feng-Lei}\
  \bibnamefont {Hong}}, \ and\ \bibinfo {author} {\bibfnamefont {Masayuki}\
  \bibnamefont {Katsuragawa}},\ }\bibfield  {title} {\enquote {\bibinfo {title}
  {Generation of five phase-locked harmonics by implementing a divide-by-three
  optical frequency divider},}\ }\href@noop {} {\bibfield  {journal} {\bibinfo
  {journal} {Opt. Lett.}\ }\textbf {\bibinfo {volume} {40}},\ \bibinfo {pages}
  {5802--5805} (\bibinfo {year} {2015})}\BibitemShut {NoStop}%
\bibitem [{\citenamefont {Kocsis}\ \emph {et~al.}(2011)\citenamefont {Kocsis},
  \citenamefont {Braverman}, \citenamefont {Ravets}, \citenamefont {Stevens},
  \citenamefont {Mirin}, \citenamefont {Shalm},\ and\ \citenamefont
  {Steinberg}}]{Kocsis2011}%
  \BibitemOpen
  \bibfield  {author} {\bibinfo {author} {\bibfnamefont {Sacha}\ \bibnamefont
  {Kocsis}}, \bibinfo {author} {\bibfnamefont {Boris}\ \bibnamefont
  {Braverman}}, \bibinfo {author} {\bibfnamefont {Sylvain}\ \bibnamefont
  {Ravets}}, \bibinfo {author} {\bibfnamefont {Martin~J.}\ \bibnamefont
  {Stevens}}, \bibinfo {author} {\bibfnamefont {Richard~P.}\ \bibnamefont
  {Mirin}}, \bibinfo {author} {\bibfnamefont {L.~Krister}\ \bibnamefont
  {Shalm}}, \ and\ \bibinfo {author} {\bibfnamefont {Aephraim~M.}\ \bibnamefont
  {Steinberg}},\ }\bibfield  {title} {\enquote {\bibinfo {title} {Observing the
  average trajectories of single photons in a two-slit interferometer},}\
  }\href {http://science.sciencemag.org/content/332/6034/1170.abstract}
  {\bibfield  {journal} {\bibinfo  {journal} {Science}\ }\textbf {\bibinfo
  {volume} {332}},\ \bibinfo {pages} {1170} (\bibinfo {year}
  {2011})}\BibitemShut {NoStop}%
\bibitem [{\citenamefont {Aspden}\ \emph {et~al.}(2016)\citenamefont {Aspden},
  \citenamefont {Padgett},\ and\ \citenamefont {Spalding}}]{Aspden2016}%
  \BibitemOpen
  \bibfield  {author} {\bibinfo {author} {\bibfnamefont {Reuben~S.}\
  \bibnamefont {Aspden}}, \bibinfo {author} {\bibfnamefont {Miles~J.}\
  \bibnamefont {Padgett}}, \ and\ \bibinfo {author} {\bibfnamefont
  {Gabriel~C.}\ \bibnamefont {Spalding}},\ }\bibfield  {title} {\enquote
  {\bibinfo {title} {Video recording true single-photon double-slit
  interference},}\ }\bibfield  {booktitle} {\emph {\bibinfo {booktitle}
  {American Journal of Physics}},\ }\href {https://doi.org/10.1119/1.4955173}
  {\bibfield  {journal} {\bibinfo  {journal} {Am. J. Phys.}\ }\textbf {\bibinfo
  {volume} {84}},\ \bibinfo {pages} {671--677} (\bibinfo {year}
  {2016})}\BibitemShut {NoStop}%
\bibitem [{\citenamefont {Eckstein}\ \emph {et~al.}(2011)\citenamefont
  {Eckstein}, \citenamefont {Christ}, \citenamefont {Mosley},\ and\
  \citenamefont {Silberhorn}}]{Eckstein2011}%
  \BibitemOpen
  \bibfield  {author} {\bibinfo {author} {\bibfnamefont {Andreas}\ \bibnamefont
  {Eckstein}}, \bibinfo {author} {\bibfnamefont {Andreas}\ \bibnamefont
  {Christ}}, \bibinfo {author} {\bibfnamefont {Peter~J.}\ \bibnamefont
  {Mosley}}, \ and\ \bibinfo {author} {\bibfnamefont {Christine}\ \bibnamefont
  {Silberhorn}},\ }\bibfield  {title} {\enquote {\bibinfo {title} {Highly
  efficient single-pass source of pulsed single-mode twin beams of light},}\
  }\href@noop {} {\bibfield  {journal} {\bibinfo  {journal} {Phys. Rev. Lett.}\
  }\textbf {\bibinfo {volume} {106}},\ \bibinfo {pages} {013603} (\bibinfo
  {year} {2011})}\BibitemShut {NoStop}%
\bibitem [{\citenamefont {Jin}\ \emph {et~al.}(2016)\citenamefont {Jin},
  \citenamefont {Shimizu}, \citenamefont {Fujiwara}, \citenamefont {Takeoka},
  \citenamefont {Wakabayashi}, \citenamefont {Yamashita}, \citenamefont {Miki},
  \citenamefont {Terai}, \citenamefont {Gerrits},\ and\ \citenamefont
  {Sasaki}}]{Jin2016QST}%
  \BibitemOpen
  \bibfield  {author} {\bibinfo {author} {\bibfnamefont {Rui-Bo}\ \bibnamefont
  {Jin}}, \bibinfo {author} {\bibfnamefont {Ryosuke}\ \bibnamefont {Shimizu}},
  \bibinfo {author} {\bibfnamefont {Mikio}\ \bibnamefont {Fujiwara}}, \bibinfo
  {author} {\bibfnamefont {Masahiro}\ \bibnamefont {Takeoka}}, \bibinfo
  {author} {\bibfnamefont {Ryota}\ \bibnamefont {Wakabayashi}}, \bibinfo
  {author} {\bibfnamefont {Taro}\ \bibnamefont {Yamashita}}, \bibinfo {author}
  {\bibfnamefont {Shigehito}\ \bibnamefont {Miki}}, \bibinfo {author}
  {\bibfnamefont {Hirotaka}\ \bibnamefont {Terai}}, \bibinfo {author}
  {\bibfnamefont {Thomas}\ \bibnamefont {Gerrits}}, \ and\ \bibinfo {author}
  {\bibfnamefont {Masahide}\ \bibnamefont {Sasaki}},\ }\bibfield  {title}
  {\enquote {\bibinfo {title} {Simple method of generating and distributing
  frequency-entangled qudits},}\ }\href@noop {} {\bibfield  {journal} {\bibinfo
   {journal} {Quantum Sci. Technol.}\ }\textbf {\bibinfo {volume} {1}},\
  \bibinfo {pages} {015004} (\bibinfo {year} {2016})}\BibitemShut {NoStop}%
\bibitem [{\citenamefont {Chen}\ \emph {et~al.}(2017)\citenamefont {Chen},
  \citenamefont {Bo}, \citenamefont {Niu}, \citenamefont {Xu}, \citenamefont
  {Zhang}, \citenamefont {Shapiro},\ and\ \citenamefont {Wong}}]{Chen2017}%
  \BibitemOpen
  \bibfield  {author} {\bibinfo {author} {\bibfnamefont {Changchen}\
  \bibnamefont {Chen}}, \bibinfo {author} {\bibfnamefont {Cao}\ \bibnamefont
  {Bo}}, \bibinfo {author} {\bibfnamefont {Murphy~Yuezhen}\ \bibnamefont
  {Niu}}, \bibinfo {author} {\bibfnamefont {Feihu}\ \bibnamefont {Xu}},
  \bibinfo {author} {\bibfnamefont {Zheshen}\ \bibnamefont {Zhang}}, \bibinfo
  {author} {\bibfnamefont {Jeffrey~H.}\ \bibnamefont {Shapiro}}, \ and\
  \bibinfo {author} {\bibfnamefont {Franco N.~C.}\ \bibnamefont {Wong}},\
  }\bibfield  {title} {\enquote {\bibinfo {title} {Efficient generation and
  characterization of spectrally factorable biphotons},}\ }\href@noop {}
  {\bibfield  {journal} {\bibinfo  {journal} {Opt. Express}\ }\textbf {\bibinfo
  {volume} {25}},\ \bibinfo {pages} {7300--7312} (\bibinfo {year}
  {2017})}\BibitemShut {NoStop}%
\bibitem [{\citenamefont {Jin}\ \emph {et~al.}(2017)\citenamefont {Jin},
  \citenamefont {Chen}, \citenamefont {Jing}, \citenamefont {Ren},
  \citenamefont {Zhao}, \citenamefont {Shimizu},\ and\ \citenamefont
  {Lu}}]{Jin2017PRA}%
  \BibitemOpen
  \bibfield  {author} {\bibinfo {author} {\bibfnamefont {Rui-Bo}\ \bibnamefont
  {Jin}}, \bibinfo {author} {\bibfnamefont {Guo-Qun}\ \bibnamefont {Chen}},
  \bibinfo {author} {\bibfnamefont {Hui}\ \bibnamefont {Jing}}, \bibinfo
  {author} {\bibfnamefont {Changliang}\ \bibnamefont {Ren}}, \bibinfo {author}
  {\bibfnamefont {Pei}\ \bibnamefont {Zhao}}, \bibinfo {author} {\bibfnamefont
  {Ryosuke}\ \bibnamefont {Shimizu}}, \ and\ \bibinfo {author} {\bibfnamefont
  {Pei-Xiang}\ \bibnamefont {Lu}},\ }\bibfield  {title} {\enquote {\bibinfo
  {title} {Monotonic quantum-to-classical transition enabled by positively
  correlated biphotons},}\ }\href@noop {} {\bibfield  {journal} {\bibinfo
  {journal} {Phys. Rev. A}\ }\textbf {\bibinfo {volume} {95}},\ \bibinfo
  {pages} {062341} (\bibinfo {year} {2017})}\BibitemShut {NoStop}%
\bibitem [{\citenamefont {Burnham}\ and\ \citenamefont
  {Weinberg}(1970)}]{Burnham1970}%
  \BibitemOpen
  \bibfield  {author} {\bibinfo {author} {\bibfnamefont {David~C.}\
  \bibnamefont {Burnham}}\ and\ \bibinfo {author} {\bibfnamefont {Donald~L.}\
  \bibnamefont {Weinberg}},\ }\bibfield  {title} {\enquote {\bibinfo {title}
  {Observation of simultaneity in parametric production of optical photon
  pairs},}\ }\href {\doibase 10.1103/PhysRevLett.25.84} {\bibfield  {journal}
  {\bibinfo  {journal} {Phys. Rev. Lett.}\ }\textbf {\bibinfo {volume} {25}},\
  \bibinfo {pages} {84--87} (\bibinfo {year} {1970})}\BibitemShut {NoStop}%
\bibitem [{\citenamefont {Hong}\ \emph {et~al.}(1987)\citenamefont {Hong},
  \citenamefont {Ou},\ and\ \citenamefont {Mandel}}]{Hong1987}%
  \BibitemOpen
  \bibfield  {author} {\bibinfo {author} {\bibfnamefont {C.~K.}\ \bibnamefont
  {Hong}}, \bibinfo {author} {\bibfnamefont {Z.~Y.}\ \bibnamefont {Ou}}, \ and\
  \bibinfo {author} {\bibfnamefont {L.}~\bibnamefont {Mandel}},\ }\bibfield
  {title} {\enquote {\bibinfo {title} {Measurement of subpicosecond time
  intervals between two photons by interference},}\ }\href@noop {} {\bibfield
  {journal} {\bibinfo  {journal} {Phys. Rev. Lett.}\ }\textbf {\bibinfo
  {volume} {59}},\ \bibinfo {pages} {2044--2046} (\bibinfo {year}
  {1987})}\BibitemShut {NoStop}%
\bibitem [{\citenamefont {Larchuk}\ \emph {et~al.}(1993)\citenamefont
  {Larchuk}, \citenamefont {Campos}, \citenamefont {Rarity}, \citenamefont
  {Tapster}, \citenamefont {Jakeman}, \citenamefont {Saleh},\ and\
  \citenamefont {Teich}}]{Larchuk1993}%
  \BibitemOpen
  \bibfield  {author} {\bibinfo {author} {\bibfnamefont {T.~S.}\ \bibnamefont
  {Larchuk}}, \bibinfo {author} {\bibfnamefont {R.~A.}\ \bibnamefont {Campos}},
  \bibinfo {author} {\bibfnamefont {J.~G.}\ \bibnamefont {Rarity}}, \bibinfo
  {author} {\bibfnamefont {P.~R.}\ \bibnamefont {Tapster}}, \bibinfo {author}
  {\bibfnamefont {E.}~\bibnamefont {Jakeman}}, \bibinfo {author} {\bibfnamefont
  {B.~E.~A.}\ \bibnamefont {Saleh}}, \ and\ \bibinfo {author} {\bibfnamefont
  {M.~C.}\ \bibnamefont {Teich}},\ }\bibfield  {title} {\enquote {\bibinfo
  {title} {Interfering entangled photons of different colors},}\ }\href@noop {}
  {\bibfield  {journal} {\bibinfo  {journal} {Phys. Rev. Lett.}\ }\textbf
  {\bibinfo {volume} {70}},\ \bibinfo {pages} {1603--1606} (\bibinfo {year}
  {1993})}\BibitemShut {NoStop}%
\bibitem [{\citenamefont {Shih}\ \emph {et~al.}(1994)\citenamefont {Shih},
  \citenamefont {Sergienko}, \citenamefont {Rubin}, \citenamefont {Kiess},\
  and\ \citenamefont {Alley}}]{Shih1994}%
  \BibitemOpen
  \bibfield  {author} {\bibinfo {author} {\bibfnamefont {Y.~H.}\ \bibnamefont
  {Shih}}, \bibinfo {author} {\bibfnamefont {A.~V.}\ \bibnamefont {Sergienko}},
  \bibinfo {author} {\bibfnamefont {M.~H.}\ \bibnamefont {Rubin}}, \bibinfo
  {author} {\bibfnamefont {T.~E.}\ \bibnamefont {Kiess}}, \ and\ \bibinfo
  {author} {\bibfnamefont {C.~O.}\ \bibnamefont {Alley}},\ }\bibfield  {title}
  {\enquote {\bibinfo {title} {Two-photon interference in a standard
  mach-zehnder interferometer},}\ }\href@noop {} {\bibfield  {journal}
  {\bibinfo  {journal} {Phys. Rev. A}\ }\textbf {\bibinfo {volume} {49}},\
  \bibinfo {pages} {4243--4246} (\bibinfo {year} {1994})}\BibitemShut {NoStop}%
\bibitem [{\citenamefont {Giovannetti}\ \emph {et~al.}(2002)\citenamefont
  {Giovannetti}, \citenamefont {Maccone}, \citenamefont {Shapiro},\ and\
  \citenamefont {Wong}}]{Giovannetti2002}%
  \BibitemOpen
  \bibfield  {author} {\bibinfo {author} {\bibfnamefont {Vittorio}\
  \bibnamefont {Giovannetti}}, \bibinfo {author} {\bibfnamefont {Lorenzo}\
  \bibnamefont {Maccone}}, \bibinfo {author} {\bibfnamefont {Jeffrey~H.}\
  \bibnamefont {Shapiro}}, \ and\ \bibinfo {author} {\bibfnamefont {Franco
  N.~C.}\ \bibnamefont {Wong}},\ }\bibfield  {title} {\enquote {\bibinfo
  {title} {Generating entangled two-photon states with coincident
  frequencies},}\ }\href@noop {} {\bibfield  {journal} {\bibinfo  {journal}
  {Phys. Rev. Lett.}\ }\textbf {\bibinfo {volume} {88}},\ \bibinfo {pages}
  {183602} (\bibinfo {year} {2002})}\BibitemShut {NoStop}%
\bibitem [{\citenamefont {Jin}\ and\ \citenamefont
  {Shimizu}(2018)}]{Jin2018Optica}%
  \BibitemOpen
  \bibfield  {author} {\bibinfo {author} {\bibfnamefont {Rui-Bo}\ \bibnamefont
  {Jin}}\ and\ \bibinfo {author} {\bibfnamefont {Ryosuke}\ \bibnamefont
  {Shimizu}},\ }\bibfield  {title} {\enquote {\bibinfo {title} {Extended
  {Wiener-Khinchin} theorem for quantum spectral analysis},}\ }\href@noop {}
  {\bibfield  {journal} {\bibinfo  {journal} {Optica}\ }\textbf {\bibinfo
  {volume} {5}},\ \bibinfo {pages} {93--98} (\bibinfo {year}
  {2018})}\BibitemShut {NoStop}%
\bibitem [{\citenamefont {Kim}\ and\ \citenamefont {Grice}(2005)}]{Kim2005}%
  \BibitemOpen
  \bibfield  {author} {\bibinfo {author} {\bibfnamefont {Yoon-Ho}\ \bibnamefont
  {Kim}}\ and\ \bibinfo {author} {\bibfnamefont {Warren~P.}\ \bibnamefont
  {Grice}},\ }\bibfield  {title} {\enquote {\bibinfo {title} {{easurement of
  the spectral properties of the two-photon state generated via type II
  spontaneous parametric downconversion}},}\ }\bibfield  {booktitle} {\emph
  {\bibinfo {booktitle} {Optics Letters}},\ }\href
  {http://ol.osa.org/abstract.cfm?URI=ol-30-8-908} {\bibfield  {journal}
  {\bibinfo  {journal} {Opt. Lett.}\ }\textbf {\bibinfo {volume} {30}},\
  \bibinfo {pages} {908--910} (\bibinfo {year} {2005})}\BibitemShut {NoStop}%
\bibitem [{\citenamefont {Shimizu}\ and\ \citenamefont
  {Edamatsu}(2009)}]{Shimizu2009}%
  \BibitemOpen
  \bibfield  {author} {\bibinfo {author} {\bibfnamefont {Ryosuke}\ \bibnamefont
  {Shimizu}}\ and\ \bibinfo {author} {\bibfnamefont {Keiichi}\ \bibnamefont
  {Edamatsu}},\ }\bibfield  {title} {\enquote {\bibinfo {title} {High-flux and
  broadband biphoton sources with controlled frequency entanglement},}\
  }\href@noop {} {\bibfield  {journal} {\bibinfo  {journal} {Opt. Express}\
  }\textbf {\bibinfo {volume} {17}},\ \bibinfo {pages} {16385--16393} (\bibinfo
  {year} {2009})}\BibitemShut {NoStop}%
\bibitem [{\citenamefont {Avenhaus}\ \emph {et~al.}(2009)\citenamefont
  {Avenhaus}, \citenamefont {Eckstein}, \citenamefont {Mosley},\ and\
  \citenamefont {Silberhorn}}]{Avenhaus2009}%
  \BibitemOpen
  \bibfield  {author} {\bibinfo {author} {\bibfnamefont {Malte}\ \bibnamefont
  {Avenhaus}}, \bibinfo {author} {\bibfnamefont {Andreas}\ \bibnamefont
  {Eckstein}}, \bibinfo {author} {\bibfnamefont {Peter~J.}\ \bibnamefont
  {Mosley}}, \ and\ \bibinfo {author} {\bibfnamefont {Christine}\ \bibnamefont
  {Silberhorn}},\ }\bibfield  {title} {\enquote {\bibinfo {title}
  {Fiber-assisted single-photon spectrograph},}\ }\href@noop {} {\bibfield
  {journal} {\bibinfo  {journal} {Opt. Lett.}\ }\textbf {\bibinfo {volume}
  {34}},\ \bibinfo {pages} {2873--2875} (\bibinfo {year} {2009})}\BibitemShut
  {NoStop}%
\bibitem [{\citenamefont {Fang}\ \emph {et~al.}(2014)\citenamefont {Fang},
  \citenamefont {Cohen}, \citenamefont {Liscidini}, \citenamefont {Sipe},\ and\
  \citenamefont {Lorenz}}]{Fang2014}%
  \BibitemOpen
  \bibfield  {author} {\bibinfo {author} {\bibfnamefont {Bin}\ \bibnamefont
  {Fang}}, \bibinfo {author} {\bibfnamefont {Offir}\ \bibnamefont {Cohen}},
  \bibinfo {author} {\bibfnamefont {Marco}\ \bibnamefont {Liscidini}}, \bibinfo
  {author} {\bibfnamefont {John~E.}\ \bibnamefont {Sipe}}, \ and\ \bibinfo
  {author} {\bibfnamefont {Virginia~O.}\ \bibnamefont {Lorenz}},\ }\bibfield
  {title} {\enquote {\bibinfo {title} {{Fast and highly resolved capture of the
  joint spectral density of photon pairs}},}\ }\href@noop {} {\bibfield
  {journal} {\bibinfo  {journal} {Optica}\ }\textbf {\bibinfo {volume} {1}},\
  \bibinfo {pages} {281} (\bibinfo {year} {2014})}\BibitemShut {NoStop}%
\bibitem [{\citenamefont {Allgaier}\ \emph {et~al.}(2017)\citenamefont
  {Allgaier}, \citenamefont {Vigh}, \citenamefont {Ansari}, \citenamefont
  {Eigner}, \citenamefont {Quiring}, \citenamefont {Ricken}, \citenamefont
  {Brecht},\ and\ \citenamefont {Silberhorn}}]{Allgaier2017}%
  \BibitemOpen
  \bibfield  {author} {\bibinfo {author} {\bibfnamefont {Markus}\ \bibnamefont
  {Allgaier}}, \bibinfo {author} {\bibfnamefont {Gesche}\ \bibnamefont {Vigh}},
  \bibinfo {author} {\bibfnamefont {Vahid}\ \bibnamefont {Ansari}}, \bibinfo
  {author} {\bibfnamefont {Christof}\ \bibnamefont {Eigner}}, \bibinfo {author}
  {\bibfnamefont {Viktor}\ \bibnamefont {Quiring}}, \bibinfo {author}
  {\bibfnamefont {Raimund}\ \bibnamefont {Ricken}}, \bibinfo {author}
  {\bibfnamefont {Benjamin}\ \bibnamefont {Brecht}}, \ and\ \bibinfo {author}
  {\bibfnamefont {Christine}\ \bibnamefont {Silberhorn}},\ }\bibfield  {title}
  {\enquote {\bibinfo {title} {Fast time-domain measurements on telecom single
  photons},}\ }\href@noop {} {\bibfield  {journal} {\bibinfo  {journal}
  {arXiv:1702.03240}\ } (\bibinfo {year} {2017})}\BibitemShut {NoStop}%
\bibitem [{\citenamefont {Kuzucu}\ \emph
  {et~al.}(2008{\natexlab{a}})\citenamefont {Kuzucu}, \citenamefont {Wong},
  \citenamefont {Kurimura},\ and\ \citenamefont {Tovstonog}}]{Kuzucu2008PRL}%
  \BibitemOpen
  \bibfield  {author} {\bibinfo {author} {\bibfnamefont {Onur}\ \bibnamefont
  {Kuzucu}}, \bibinfo {author} {\bibfnamefont {Franco N.~C.}\ \bibnamefont
  {Wong}}, \bibinfo {author} {\bibfnamefont {Sunao}\ \bibnamefont {Kurimura}},
  \ and\ \bibinfo {author} {\bibfnamefont {Sergey}\ \bibnamefont {Tovstonog}},\
  }\bibfield  {title} {\enquote {\bibinfo {title} {Joint temporal density
  measurements for two-photon state characterization},}\ }\href@noop {}
  {\bibfield  {journal} {\bibinfo  {journal} {Phys. Rev. Lett.}\ }\textbf
  {\bibinfo {volume} {101}},\ \bibinfo {pages} {153602} (\bibinfo {year}
  {2008}{\natexlab{a}})}\BibitemShut {NoStop}%
\bibitem [{\citenamefont {Cho}\ \emph {et~al.}(2014)\citenamefont {Cho},
  \citenamefont {Park}, \citenamefont {Lee},\ and\ \citenamefont
  {Kim}}]{Cho2014}%
  \BibitemOpen
  \bibfield  {author} {\bibinfo {author} {\bibfnamefont {Young-Wook}\
  \bibnamefont {Cho}}, \bibinfo {author} {\bibfnamefont {Kwang-Kyoon}\
  \bibnamefont {Park}}, \bibinfo {author} {\bibfnamefont {Jong-Chan}\
  \bibnamefont {Lee}}, \ and\ \bibinfo {author} {\bibfnamefont {Yoon-Ho}\
  \bibnamefont {Kim}},\ }\bibfield  {title} {\enquote {\bibinfo {title}
  {Engineering frequency-time quantum correlation of narrow-band biphotons from
  cold atoms},}\ }\href@noop {} {\bibfield  {journal} {\bibinfo  {journal}
  {Phys. Rev. Lett.}\ }\textbf {\bibinfo {volume} {113}},\ \bibinfo {pages}
  {063602} (\bibinfo {year} {2014})}\BibitemShut {NoStop}%
\bibitem [{\citenamefont {K{\"o}nig}\ and\ \citenamefont
  {Wong}(2004)}]{Konig2004}%
  \BibitemOpen
  \bibfield  {author} {\bibinfo {author} {\bibfnamefont {Friedrich}\
  \bibnamefont {K{\"o}nig}}\ and\ \bibinfo {author} {\bibfnamefont {Franco
  N.~C.}\ \bibnamefont {Wong}},\ }\bibfield  {title} {\enquote {\bibinfo
  {title} {Extended phase matching of second-harmonic generation in
  periodically poled {KTiOPO}$_4$ with zero group-velocity mismatch},}\
  }\href@noop {} {\bibfield  {journal} {\bibinfo  {journal} {Appl. Phys.
  Lett.}\ }\textbf {\bibinfo {volume} {84}},\ \bibinfo {pages} {1644} (\bibinfo
  {year} {2004})}\BibitemShut {NoStop}%
\bibitem [{\citenamefont {Jin}\ \emph {et~al.}(2013)\citenamefont {Jin},
  \citenamefont {Shimizu}, \citenamefont {Wakui}, \citenamefont {Benichi},\
  and\ \citenamefont {Sasaki}}]{Jin2013OE}%
  \BibitemOpen
  \bibfield  {author} {\bibinfo {author} {\bibfnamefont {Rui-Bo}\ \bibnamefont
  {Jin}}, \bibinfo {author} {\bibfnamefont {Ryosuke}\ \bibnamefont {Shimizu}},
  \bibinfo {author} {\bibfnamefont {Kentaro}\ \bibnamefont {Wakui}}, \bibinfo
  {author} {\bibfnamefont {Hugo}\ \bibnamefont {Benichi}}, \ and\ \bibinfo
  {author} {\bibfnamefont {Masahide}\ \bibnamefont {Sasaki}},\ }\bibfield
  {title} {\enquote {\bibinfo {title} {Widely tunable single photon source with
  high purity at telecom wavelength},}\ }\href@noop {} {\bibfield  {journal}
  {\bibinfo  {journal} {Opt. Express}\ }\textbf {\bibinfo {volume} {21}},\
  \bibinfo {pages} {10659--10666} (\bibinfo {year} {2013})}\BibitemShut
  {NoStop}%
\bibitem [{\citenamefont {Kuzucu}\ \emph
  {et~al.}(2008{\natexlab{b}})\citenamefont {Kuzucu}, \citenamefont {Wong},
  \citenamefont {Kurimura},\ and\ \citenamefont {Tovstonog}}]{Kuzucu2008OL}%
  \BibitemOpen
  \bibfield  {author} {\bibinfo {author} {\bibfnamefont {Onur}\ \bibnamefont
  {Kuzucu}}, \bibinfo {author} {\bibfnamefont {Franco N.~C.}\ \bibnamefont
  {Wong}}, \bibinfo {author} {\bibfnamefont {Sunao}\ \bibnamefont {Kurimura}},
  \ and\ \bibinfo {author} {\bibfnamefont {Sergey}\ \bibnamefont {Tovstonog}},\
  }\bibfield  {title} {\enquote {\bibinfo {title} {Time-resolved single-photon
  detection by femtosecond upconversion},}\ }\href@noop {} {\bibfield
  {journal} {\bibinfo  {journal} {Opt. Lett.}\ }\textbf {\bibinfo {volume}
  {33}},\ \bibinfo {pages} {2257--2259} (\bibinfo {year}
  {2008}{\natexlab{b}})}\BibitemShut {NoStop}%
\bibitem [{\citenamefont {Nielsen}\ and\ \citenamefont
  {Chuang}(2000)}]{Nielsen2000}%
  \BibitemOpen
  \bibfield  {author} {\bibinfo {author} {\bibfnamefont {Michael~A.}\
  \bibnamefont {Nielsen}}\ and\ \bibinfo {author} {\bibfnamefont {Isaac~L.}\
  \bibnamefont {Chuang}},\ }\href@noop {} {\emph {\bibinfo {title} {Quantum
  Computation and Quantum Information}}}\ (\bibinfo  {publisher} {Cambridge
  University Press},\ \bibinfo {year} {2000})\BibitemShut {NoStop}%
\bibitem [{\citenamefont {Schweigler}\ \emph {et~al.}(2017)\citenamefont
  {Schweigler}, \citenamefont {Kasper}, \citenamefont {Erne}, \citenamefont
  {Mazets}, \citenamefont {Rauer}, \citenamefont {Cataldini}, \citenamefont
  {Langen}, \citenamefont {Gasenzer}, \citenamefont {Berges},\ and\
  \citenamefont {Schmiedmayer}}]{Schweigler2017}%
  \BibitemOpen
  \bibfield  {author} {\bibinfo {author} {\bibfnamefont {Thomas}\ \bibnamefont
  {Schweigler}}, \bibinfo {author} {\bibfnamefont {Valentin}\ \bibnamefont
  {Kasper}}, \bibinfo {author} {\bibfnamefont {Sebastian}\ \bibnamefont
  {Erne}}, \bibinfo {author} {\bibfnamefont {Igor}\ \bibnamefont {Mazets}},
  \bibinfo {author} {\bibfnamefont {Bernhard}\ \bibnamefont {Rauer}}, \bibinfo
  {author} {\bibfnamefont {Federica}\ \bibnamefont {Cataldini}}, \bibinfo
  {author} {\bibfnamefont {Tim}\ \bibnamefont {Langen}}, \bibinfo {author}
  {\bibfnamefont {Thomas}\ \bibnamefont {Gasenzer}}, \bibinfo {author}
  {\bibfnamefont {J{\"{u}}rgen}\ \bibnamefont {Berges}}, \ and\ \bibinfo
  {author} {\bibfnamefont {J{\"{o}}rg}\ \bibnamefont {Schmiedmayer}},\
  }\bibfield  {title} {\enquote {\bibinfo {title} {{Experimental
  characterization of a quantum many-body system via higher-order
  correlations}},}\ }\href@noop {} {\bibfield  {journal} {\bibinfo  {journal}
  {Nature}\ }\textbf {\bibinfo {volume} {545}},\ \bibinfo {pages} {323--326}
  (\bibinfo {year} {2017})}\BibitemShut {NoStop}%
\bibitem [{\citenamefont {Bisht}\ and\ \citenamefont
  {Shimizu}(2015)}]{Bisht2015}%
  \BibitemOpen
  \bibfield  {author} {\bibinfo {author} {\bibfnamefont {Nandan~S.}\
  \bibnamefont {Bisht}}\ and\ \bibinfo {author} {\bibfnamefont {Ryosuke}\
  \bibnamefont {Shimizu}},\ }\bibfield  {title} {\enquote {\bibinfo {title}
  {{Spectral properties of broadband biphotons generated from PPMgSLT under a
  type-II phase-matching condition}},}\ }\href {\doibase
  10.1364/JOSAB.32.000550} {\bibfield  {journal} {\bibinfo  {journal} {J. Opt.
  Soc. Am. B}\ }\textbf {\bibinfo {volume} {32}},\ \bibinfo {pages} {550--554}
  (\bibinfo {year} {2015})}\BibitemShut {NoStop}%
\bibitem [{\citenamefont {Evans}\ \emph {et~al.}(2010)\citenamefont {Evans},
  \citenamefont {Bennink}, \citenamefont {Grice}, \citenamefont {Humble},\ and\
  \citenamefont {Schaake}}]{Evans2010}%
  \BibitemOpen
  \bibfield  {author} {\bibinfo {author} {\bibfnamefont {P.~G.}\ \bibnamefont
  {Evans}}, \bibinfo {author} {\bibfnamefont {R.~S.}\ \bibnamefont {Bennink}},
  \bibinfo {author} {\bibfnamefont {W.~P.}\ \bibnamefont {Grice}}, \bibinfo
  {author} {\bibfnamefont {T.~S.}\ \bibnamefont {Humble}}, \ and\ \bibinfo
  {author} {\bibfnamefont {J.}~\bibnamefont {Schaake}},\ }\bibfield  {title}
  {\enquote {\bibinfo {title} {Bright source of spectrally uncorrelated
  polarization-entangled photons with nearly single-mode emission},}\
  }\href@noop {} {\bibfield  {journal} {\bibinfo  {journal} {Phys. Rev. Lett.}\
  }\textbf {\bibinfo {volume} {105}},\ \bibinfo {pages} {253601} (\bibinfo
  {year} {2010})}\BibitemShut {NoStop}%
\bibitem [{\citenamefont {Gerrits}\ \emph {et~al.}(2011)\citenamefont
  {Gerrits}, \citenamefont {Stevens}, \citenamefont {Baek}, \citenamefont
  {Calkins}, \citenamefont {Lita}, \citenamefont {Glancy}, \citenamefont
  {Knill}, \citenamefont {Nam}, \citenamefont {Mirin}, \citenamefont
  {Hadfield}, \citenamefont {Bennink}, \citenamefont {Grice}, \citenamefont
  {Dorenbos}, \citenamefont {Zijlstra}, \citenamefont {Klapwijk},\ and\
  \citenamefont {Zwiller}}]{Gerrits2011}%
  \BibitemOpen
  \bibfield  {author} {\bibinfo {author} {\bibfnamefont {Thomas}\ \bibnamefont
  {Gerrits}}, \bibinfo {author} {\bibfnamefont {Martin~J.}\ \bibnamefont
  {Stevens}}, \bibinfo {author} {\bibfnamefont {Burm}\ \bibnamefont {Baek}},
  \bibinfo {author} {\bibfnamefont {Brice}\ \bibnamefont {Calkins}}, \bibinfo
  {author} {\bibfnamefont {Adriana}\ \bibnamefont {Lita}}, \bibinfo {author}
  {\bibfnamefont {Scott}\ \bibnamefont {Glancy}}, \bibinfo {author}
  {\bibfnamefont {Emanuel}\ \bibnamefont {Knill}}, \bibinfo {author}
  {\bibfnamefont {Sae~Woo}\ \bibnamefont {Nam}}, \bibinfo {author}
  {\bibfnamefont {Richard~P.}\ \bibnamefont {Mirin}}, \bibinfo {author}
  {\bibfnamefont {Robert~H.}\ \bibnamefont {Hadfield}}, \bibinfo {author}
  {\bibfnamefont {Ryan~S.}\ \bibnamefont {Bennink}}, \bibinfo {author}
  {\bibfnamefont {Warren~P.}\ \bibnamefont {Grice}}, \bibinfo {author}
  {\bibfnamefont {Sander}\ \bibnamefont {Dorenbos}}, \bibinfo {author}
  {\bibfnamefont {Tony}\ \bibnamefont {Zijlstra}}, \bibinfo {author}
  {\bibfnamefont {Teun}\ \bibnamefont {Klapwijk}}, \ and\ \bibinfo {author}
  {\bibfnamefont {Val}\ \bibnamefont {Zwiller}},\ }\bibfield  {title} {\enquote
  {\bibinfo {title} {Generation of degenerate, factorizable, pulsed squeezed
  light at telecom wavelengths},}\ }\href@noop {} {\bibfield  {journal}
  {\bibinfo  {journal} {Opt. Express}\ }\textbf {\bibinfo {volume} {19}},\
  \bibinfo {pages} {24434--24447} (\bibinfo {year} {2011})}\BibitemShut
  {NoStop}%
\end{thebibliography}
\end{document}